\documentclass[extra,mreferee]{gji} %review version
\usepackage{timet}
\usepackage{amsmath,amssymb} 
\usepackage{graphicx}
\usepackage{lineno}

\usepackage{ulem}
\usepackage{color}

\title[
Paradox of Curved Faults Revisited 
]
  {Paradox of Modeling Curved Faults Revisited with General Non-Hypersingular Stress Green's Functions}
\author[D. Sato, P. Romanet, \& R. Ando]{Dye SK Sato$^1$, Pierre Romanet$^2$, and Ryosuke Ando$^2$
\\
  $^1$ Disaster Prevention Research Institute, Kyoto University, Gokasho, Uji, Kyoto 611-0011, Japan
\\
  $^2$ The University of Tokyo, 7-3-1 Hongo, Bunkyo-ku, Tokyo 113-8654, Japan}
\date{Received 20xx Xxxx xx; in original form 20xx Xxxx xx}
\pagerange{\pageref{firstpage}--\pageref{lastpage}}
\volume{xxx}
\pubyear{xxxx}

\begin{document}

\label{firstpage}

\maketitle

%\linenumbers
\begin{summary}
% This guide is for authors who are preparing papers for
% \textit{Geophysical Journal International} using the
%\LaTeXe\ document preparation system and the GJI class file.
In a dislocation problem, a paradoxical discordance is known to occur 
between an original smooth curve and an infinitesimally discretized curve. 
To solve this paradox, we have investigated a non-hypersingular expression for the integral kernel (called the stress Green's function) which describes the stress field caused by the displacement discontinuity. 
We first develop a compact alternative expression of the non-hypersingular stress Green's function for general two- and three-dimensional infinite homogeneous elastic media. 
We next compute the stress Green's functions on a curved fault and revisit the paradox. 
We find that previously obtained non-hypersingular stress Green's functions are incorrect for curved faults, and that smooth and infinitesimally segmented faults are equivalent. Their compatibility bridges the gap between analytical methods featuring curved faults and numerical methods using subdivided flat patches.
%verifies macroscopic kinematics of neglecting fine scales in a dislocation problem and the most importantly the numerics of discretizing non-planar fault geometries.}
\end{summary}


\section{
Introduction
}
Non-planar fault geometries, such as bends, branches, and segments, have been considered to affect the rupture processes of the earthquakes~\citep{scholz2019mechanics}. 
The rupture processes on non-planar faults are difficult to solve analytically, and sometimes even numerically. A fundamental tool for investigating such phenomena is the established solution of internal deformation due to a piecewise discrete slip on a flat fault element~\citep[e.g.,][]{okada1992internal,aochi2000spontaneous}, which is obtained analytically from the integral kernel (Green's function) for elastic media. 
By discretizing the intractable general fault geometry into the amenable flat fault elements~\citep{cochard1994dynamic,tada2001dynamic,aochi2002three,tada2006stress}, analytical results have been successful in modelling the earthquake rupture of the complex fault geometries~\citep{rice1993spatio,aochi2002three,kame2003effects,ando2018dynamic}.

In such fault-modeling discretization, it is assumed that the discretized solution converges to the original un-discretized one in the limit of the infinitesimally fine elements \citep{tada1996paradox}. However, \citet{tada1996paradox} reported that discretized solutions for problems involving smoothly curved faults do not converge to the original un-discretized ones, even in the limit of infinitesimally fine elements. 
They further pointed out that this inconsistency extends to the previous analytical formulations, such as that of \citet{jeyakumaran1994curved}, as well as to their own results.
Although \citet{tada1996paradox} first recognized this problem in two-dimensional cases, 
 \citet{aochi2000spontaneous} and \citet{tada2000non} reported that this inconsistency 
persists in three-dimensional modeling as well. This inconsistency has been considered to be a paradox of elastic fault-modeling~\citep{tada1996paradox,aochi2000spontaneous}, called the ``paradox of smooth and abrupt bends''~\citep{tada1996paradox}.

The paradox of smooth and abrupt bends posed the issue of whether a fully smooth or a finely discretized/segmented fault is appropriate for modeling a real curved fault~\citep{aki2002quantitative,duan2005multicycle,kase2006spontaneous}. By analogy to a natural fault that is segmented on fine scales, some investigators have considered that a discretized fault may be more appropriate, at least for purely elastic problems~\citep{duan2005multicycle,kase2006spontaneous}. \citet{kase2006spontaneous} also reported that the solution obtained by using discretized boundary elements is well supported by the finite-element modeling approach, as done e.g., by \citet[][]{oglesby20031999}, but it also uses the discretized fault segments implicitly and so cannot verify the discretization of boundary geometry. In addition, the dynamic rupture problem on a kinked fault becomes ill-posed, due to the ambiguity of the slip direction at a corner~\citep{adda2008seismic}. 
Despite the fact that a discretized fault was first introduced as a tractable approximation to a smooth curve, researchers have found it necessary to be conscious of the case to which their numerical modelling corresponds, so as not to misinterpret numerical results~\citep{tada1996paradox}.

%Hence the infinitesimally finely discretized fault can be ill-posed everywhere, unless it is treated as a mere numerical discretization of the smooth curve. 
%Besides, although above studies are for the purely elastic media, it is also unknown whether the paradox of curved faults is for the dislocation problem in the purely elastic media or a paradox remaining in the continuum mechanics of general constitutive properties such as the off-fault damage rheology.

%For bridging the gap between smooth and discretized faults, 
In this paper, we study the paradox of smooth and abrupt bends from both analytical and numerical viewpoints in order to investigate the adequacy of the choice between smooth and segmented faults. 
In previous studies, the paradox of smooth and abrupt bends was considered by using only the analytically obtained non-hypersingular stress Green's functions, as in~\citet{jeyakumaran1994curved,tada1996boundary,tada1997non,tada2000non}. 
Because of the length of the expressions obtained in those studies, the cause of the paradox of smooth and abrupt bends does not appear clearly or become disentangled in a unified manner. 
By investigating the derivation of a compact expression of the stress Green's function--which we find to be equivalent to the expression obtained by \citet{bonnet1999boundary}--and by comparing the previously obtained non-hypersingular expressions numerically, we show that there is in fact no paradox.

This paper is organized as follows. First, we present the definition of the problem. Second, we investigate the derivation of the non-hypersingular stress Green's functions for both two- and three-dimensional problems. Third, we examine the paradox of smooth and abrupt bends in various numerical ways. Finally, we provide an intuitive explanation of the results, which connect to the central topic in the companion manuscript to this paper (Romanet, Sato, and Ando). 

\section{
Definition of the Problem
}
The problem we address is to obtain the non-hypersingular stress Green's function, a boundary integral equation that describes the stress field throughout a medium due to displacement discontinuities on the fault. 
%The large deformation problem is out of the scope in this study as added in \S\ref{sec:33}.
We consider a homogeneous, elastic medium filling the infinite space with the absolute stress in static equilibrium and without any single forces. 
We do not assume the medium to be isotropic.
%Isotropy of the medium is not assumed. 
We assume small displacements, and express the fault as the buried boundary that constitutes the interface between two sufficiently adjacent faces~\citep[p38]{aki2002quantitative}. 
The scope of our derivation includes those of the previous studies that have obtained non-hypersingular stress Green's functions for isotropic elasticity~\citep[e.g.,][]{tada1997non,tada2000non}. We consider the three-dimensional dynamic case, as it reduces to the other cases (static or two-dimensional) in certain limits.

%We begin the derivation with the equation of motion (\S\ref{sec:21}).  
We begin with the representation theorem (\S\ref{sec:22}), which provides the hypersingular integral equations for displacement gradient and stress Green's functions (\S\ref{sec:23}).

\subsection{Representation Theorem} \label{sec:22}
%The displacement field in the elastic medium
%giving $\sigma=\sigma_{el}$ for arbitrary frequency ranges, 
%is given through Betti's representation theorem over the entire spaces $V$ of the medium from the variables on the set of the fault areas $\Gamma(s)$ that can depend on the time $s$.

The $n$-th component $u_n({\bf x},t)$ of the displacement vector ${\bf u}({\bf x},t)$ at location ${\bf x}=(x_1,x_2,x_3)$ and time $t$ is described by the representation theorem as a function of the slip distance ${\bf\Delta u}(\boldsymbol \xi, s)$ distributed over locations $\boldsymbol \xi$ on the set of faults $\Gamma(s)$ at time $s$, where the Latin subscripts range over the set $\{1,2,3\}$. 
The representation theorem for buried faults~\citep{aki2002quantitative} is given as
\begin{eqnarray}
u_n({\bf x},t) 
&=& 
 \int^\infty_{-\infty} ds
\int_{\Gamma(s)} d\Sigma(\boldsymbol\xi)  
\Delta u_{i}(\boldsymbol\xi,s) 
\nonumber\\&&\times
\nu_{j}(\boldsymbol\xi,s) c_{ijpq} \frac{\partial G_{np}}{\partial \xi_q} ({\bf x},t;\boldsymbol\xi,s),
\label{eq:Bettihom}
\end{eqnarray}
where 
$c_{ijpq}$ is the $ijpq$ component of the elasticity tensor, 
$\Delta u_i({\boldsymbol \xi}, s)$ is the $i$-th component of ${\bf\Delta u}(\boldsymbol \xi, s)$,
and the boundary integral $\int_{\Gamma(s)} d\Sigma(\boldsymbol\xi)$ is executed over $\Gamma(s)$, and $\nu_{j}(\boldsymbol \xi,s)$ is the $j$-th component of the inward normal vector $\boldsymbol \nu(\boldsymbol \xi,s)$ [of the upper surface of $\Gamma(s)$] at the location $\boldsymbol\xi$ on $\Gamma(s)$ at the time $s$. Summation over repeated Latin subscripts is implied.
The quantity $G_{in}$ is the $in$-component of the retarded homogeneous Green's function ${\bf G}$ of the displacement field that obeys the equation of motion in a homogeneous medium filling infinite space; it describes the $i$-th component of the displacement at location ${\bf x}$ and time $t$ in responding to a delta-impulsive single force along the $n$-th direction at the location 
$\boldsymbol \xi$ and the time $s$. That is 
\begin{eqnarray}
&&
\rho\partial_t^2 G_{in}({\bf x},t;\xi,s)
\nonumber\\&=&
\partial_j^{(x)} [c_{ijpq}\partial^{(x)}_p G_{qn}({\bf x},t;\xi,s)]
+\delta_{in}\delta ({\bf x}-\boldsymbol\xi)\delta(t-s),
\label{eq:EOMofG}
\end{eqnarray}
where 
$\rho$ is the mass density of the medium, 
$\delta_{in}$ is the Kronecker delta for the set $(i,j)$, and $\delta({\bf x}-\boldsymbol\xi)$ and $\delta (t-s)$ are Dirac delta functions for the relative location ${\bf x}-\boldsymbol\xi$ and relative time $t-s$. The quantities $\partial_t:=\partial/(\partial t)$ and $\partial_j:=\partial/(\partial x_j)$, respectively, represent the partial-differentiation operators for the time $t$ and the $j$-th component $x_j$ of the position vector ${\bf x}$. 
Hereafter, when one side of an equation contains both ${\bf x}$ and $\boldsymbol \xi$, we specify ${\bf x}$ or $\boldsymbol \xi$ to execute spatial differentiations as $\partial^{(x)}_m=\partial/(\partial x_m)$ or $\partial^{(\xi)}_m=\partial/(\partial\xi_m)$.
Note that the homogeneous Green's function refers to the homogeneous solution of Eq.~(\ref{eq:EOMofG}), which satisfies boundary conditions of zero displacement $G_{in}({\bf x},t;\boldsymbol\xi,s)=0$ (the rigid boundary condition) or zero stress (strain) $\nu_{j}({\bf x},t) c_{ijkl}\partial_k^{(x)}$$G_{ln}$$({\bf x},t;\boldsymbol \xi,s)$$=0$ (the free surface condition),
as in the context of the partial differential equations; ${\bf G}$ of this study is then a homogeneous solution for the homogeneous medium of full space.

The slip distance ${\bf \Delta  u}(\boldsymbol \xi, s)$ at location $\boldsymbol \xi$ on the fault at time $s$ is defined by using the normal vector ${\boldsymbol \nu}$ at location $\boldsymbol \xi$ 
as 
\begin{eqnarray}
{\bf \Delta  u}(\boldsymbol \xi,s):=\lim_{\delta \to0}[{\bf u}(\boldsymbol\xi+\delta\boldsymbol\nu,s)-{\bf u}(\boldsymbol \xi-\delta\boldsymbol \nu,s)].
\end{eqnarray}
This represents the displacement difference between the two faces of the fault.
Below, we abbreviate the set of the faults $\Gamma$ as ``the fault'' for brevity.
At the edge of each finite-sized fault, ${\bf \Delta u}={\bf 0}$ is satisfied. 
%We use the condition ${\bf G}=0$ at infinity as the edge condition for infinitely long faults in subsequent analysis. 
$\Delta {\bf u}={\bf 0}$ is not necessarily satisfied at the edge of a boundary of infinite length, and so we will use the condition ${\bf G} = {\bf 0}$ at infinity instead of $\Delta {\bf u}={\bf 0}$ for the edges of such infinitely long faults in subsequent derivation; for a finite-sized fault of no edges (which is inevitably a seamless surface that bounds a closed space), we will use the continuity of $\Delta {\bf u}$ in the derivation.

%\subsection{Spatiotemporal Translational Symmetries and Reciprocities of Homogeneous Green's Function}\label{sec:31}
To derive the non-hypersingular stress Green's function,
we rely on the spatiotemporal symmetry of the homogeneous Green's function: %, that is Green's function $G ({\bf x}, t; \boldsymbol \xi, s)$ for the homogeneous medium covering the infinite space, 
%that it depends %is reduced to a function %${\bf G}({\bf x}-\boldsymbol \xi, t-s)$ 
%only on the relative location ${\bf x}-\boldsymbol \xi$ and the relative time $t-s$ and satisfies 
\begin{eqnarray}
\partial_m^{(x)}
{\bf G}({\bf x}-\boldsymbol \xi, t-s)
=
-\partial_ m^{(\xi)}
{\bf G}({\bf x}-\boldsymbol \xi, t-s)
\label{eq:spatialtrans}
\\
\partial_t
{\bf G}({\bf x}-\boldsymbol \xi, t-s)
=
-\partial_ s
{\bf G}({\bf x}-\boldsymbol \xi, t-s)
\label{eq:temporaltrans}
\end{eqnarray}
where $\partial_s$ represents partial differentiation with respect to the time $s$. 
These relations mean that the homogeneous Green's function depends only on the relative location ${\bf x}-\boldsymbol \xi$ and the relative time $t-s$.
%Eqs.~(\ref{eq:temporaltrans}) and (\ref{eq:spatialtrans}) play the intrinsic roles 
We emphasize that Eq.~(\ref{eq:spatialtrans}) is valid specially for a homogeneous infinite medium, independent of the isotropy and non-isotropy, and is generally not applicable to inhomogeneous or bounded media. It limits the applicability of our derivation relying on Eq.~(\ref{eq:spatialtrans}). The extension of our formulation to heterogeneous media needs an indirect approach, as mentioned in the discussion section.

In the derivation of the stress Green's function, we utilize the spatial reciprocity of Green's function,
\begin{eqnarray}
G_{in}({\bf x},t;\boldsymbol\xi,s)=G_{ni}(\boldsymbol\xi,t;{\bf x},s).
\label{eq:spatialrecipro}
\end{eqnarray}
Note that Eq.~(\ref{eq:spatialrecipro}) is valid as long as 
homogeneous boundary conditions are imposed on all the boundaries except the fault $\Gamma$~\citep[p29]{aki2002quantitative}. 
We will also recall the symmetry of the elasticity tensor:
\begin{eqnarray}
c_{ijkl}=c_{jikl}=c_{ijlk}=c_{klij}.
\label{eq:symofc}
\end{eqnarray}

%Hereafter, ${\bf G}$ is supposed to be ${\bf G}_{hom}$ (${\bf G}={\bf G}_{hom}$).
%Although the explicit form of Green's function depends on the property of the elasticity tensor and is not known for general cases, such explicit form is not required in our regularization.

%\footnote{
%$G$は相対位置${\bf x}-\boldsymbol \xi$と相対時間$t-s$のみに依存する$G_{hom} ({\bf x}-\boldsymbol \xi, t-s)$へと縮約される( $G ({\bf x},t;\xi,s)=G_{hom}({\bf x}-\boldsymbol \xi, t-s)$)。
%以降の導出では、この並進性に起因するreciprocityが本質的な役割を果たす。
%
%ここで時間$s$の偏微分を..と書いた。なお、以降の導出で具体的な式の形は要求されない。
%}

\subsection{Hypersingular Displacement-Gradient and Stress Green's Functions}\label{sec:23}

Spatial differentiation of Eq.~(\ref{eq:Bettihom}) gives the spatiotemporal distribution of the displacement gradient:
\begin{eqnarray}
\partial_m u_n({\bf x},t) &=&
\int_{-\infty}^{\infty} ds 
\int_{\Gamma(s)} d\Sigma(\boldsymbol\xi) 
\Delta u_{i}(\boldsymbol\xi,s) 
\nonumber\\&&\times
\nu_{j}(\boldsymbol\xi,s) c_{ijpq} 
\partial_m^{(x)}
\frac{\partial G_{np}}{\partial \xi_q} ({\bf x},t;\boldsymbol\xi,s).
\label{eq:displacementgradienthom}
\end{eqnarray}
This integral equation for the response of the displacement gradient to the displacement (called the ``displacement-gradient Green's function''), Eq.~(\ref{eq:displacementgradienthom}), is known to become hypersingular. That is, it cannot be evaluated as a Cauchy integral, even when the convolved variable is H\"{o}lder-continuous~\citep{koller1992modelling}. 
For example, the integral kernel in Eq.~(\ref{eq:displacementgradienthom}) contains a term proportional to $r^{-3}\delta(t-r/\beta)$ in the isotropic case, where $\beta$ is the S-wave speed, and hence the integral equation diverges.
Our aim is to make such an integral kernel integrable in the Cauchy sense (called ``regularization'') as long as it is convolved with H\"{o}lder-continuous boundary variables. For example, such boundary variables can be the spatial derivatives of the slip along the boundary (the dislocation), and not necessarily the slip itself.

%Eq.~(\ref{eq:reexofstress}) gives the stress field with 
With the constitutive law of the elasticity, 
\begin{eqnarray}
\sigma_{kl} ({\bf x},t)=c_{klmn}\partial_m u_n({\bf x},t),
\label{eq:reexofstress}
\end{eqnarray}
Eq.~(\ref{eq:displacementgradienthom}) also gives the hypersingular stress Green's function, as well as the hypersingular Green's function for the strain $(\partial_m u_n+\partial_n u_m)/2$.
Here, $\sigma_{kl} ({\bf x},t)$ denotes the $kl$ component of the stress $\boldsymbol\sigma ({\bf x},t)$ at each location ${\bf x}$ and time $t$. Note that this expression is obtained from the ordinary expression $\sigma_{kl}=c_{klmn}(\partial_mu_n+\partial_n u_m)/2$, with the symmetry $c_{klmn}=c_{klnm}$ shown in Eq.~(\ref{eq:symofc}). 
% and the definition of the symmetric strain Eq.~(\ref{eq:defofstress}) simplifies the definition of stress (Eq.~(\ref{eq:defofstress})) as where the index is changed from $ijkl$ of Eq.~(\ref{eq:defofstress}) to $klmn$ for being identified with that of Eq.~(\ref{eq:displacementgradient}).
The traction ${\bf T}({\bf x},t)$ at the location ${\bf x}$ on the fault $\Gamma$ (${\bf x}\in \Gamma$) at time $t$ is also given in the tensorial form $T_i=\sigma_{ij}\nu_j$. 
%The strain and stress Green's functions obtained from Eq.~(\ref{eq:displacementgradienthom}) are also hypersingular. 

Hereafter we assume the differentiability of the slip with respect to time and space in order to develop the non-hypersingular expressions. This holds for the conventional rock-mechanical cases assuming a smooth slip gradient (a spatial differential of the slip) and slip rate (temporal one). The assumptions concerning $\Gamma$ is discussed  when we introduce the local coordinate system in \S\ref{sec:32}.

\section{
General Forms of the Non-Hypersingular Displacement-Gradient and Stress Green's Functions
}
%The hypersingular integral equation for the displacement gradient field 
%Eq.~(\ref{eq:displacementgradienthom}) was introduced from Betti's representation theorem Eq.~(\ref{eq:Bettihom}) in \S\ref{sec:23}.
%Below, by regularizing Eq.~(\ref{eq:displacementgradienthom}),
%we derive the non-hypersingular integral equations for the displacement gradient, and the non-hypersingular stress Green's functions. 
%これによって、断層面上の滑り、あるいは滑りレートで閉じることの可能な積分方程式の表現を求めることが最終的な目標である。

We regularize Eq.~(\ref{eq:displacementgradienthom}) by following the widely adopted direct approach of regularization in real space~\citep[e.g.,][]{koller1992modelling,cochard1994dynamic,tada1997non,tada2000non}. This process can be unified by utilizing the equation of motion, Eq.~(\ref{eq:EOMofG}), %of Green's function 
and the translational symmetry %(detailed in \S\ref{sec:31}) 
of the Green's function~\citep{bonnet1999boundary}. %Other approaches are out of the scope of this study, such as that based on the variational approaches~\citep[e.g.,][]{nedelec1982integral,nishimura1989regularized,bonnet1995regularized} using Kelvin's fundamental solution, and an indirect way~\citep[e.g.,][]{bonnet1993regularization} requiring the residue integral. 
We perform the regularization by using coordinates spanned along the boundary (the local coordinates) in \S\ref{sec:32}, which provides a unified way of structuring the non-hypersingular displacement-gradient and stress Green's functions in \S\ref{sec:33}.

The following regularization of the stress Green's function does not require the discrimination between the fault $\Gamma$ and other boundaries. We therefore refer to ``the boundary'' rather than ``the fault'' in this section, unless otherwise necessary.

\subsection{
Local Coordinate System
}\label{sec:32}
%The local coordinate is detialed below.
The local coordinate system is first defined in \S\ref{sec:321}. 
We subsequently relate the coordinate values of the local coordinate system and of the global coordinate system spanned by $(x_1,x_2,x_3)$ axes in \S\ref{sec:322}. A useful equality concerning the local coordinate is introduced in \S\ref{sec:323} for regularization. 

To provide an intuitive explanation, we here suppose that the whole boundary area referred to by $\Gamma$ is on a single boundary, the geometry of which is describable by a spatiotemporal function of class $C^1$ (which allows differentiation once). That is, two arbitrary points on $\Gamma$ can be connected by a path of class $C^1$ on $\Gamma$.  
This simplification is solely for the explanatory purpose, and the relations introduced in \S\ref{sec:32} generally hold in each boundary as long as $\Gamma$ consists of multiple boundaries of class $C^1$, e.g., multiple unjointed faults of class $C^1$. Such a relation also holds on a kinked fault as long as it can be represented by connected smooth boundaries.

\subsubsection{Definition}
\label{sec:321}
\begin{figure}%[tbp]
  \includegraphics[width=80mm]{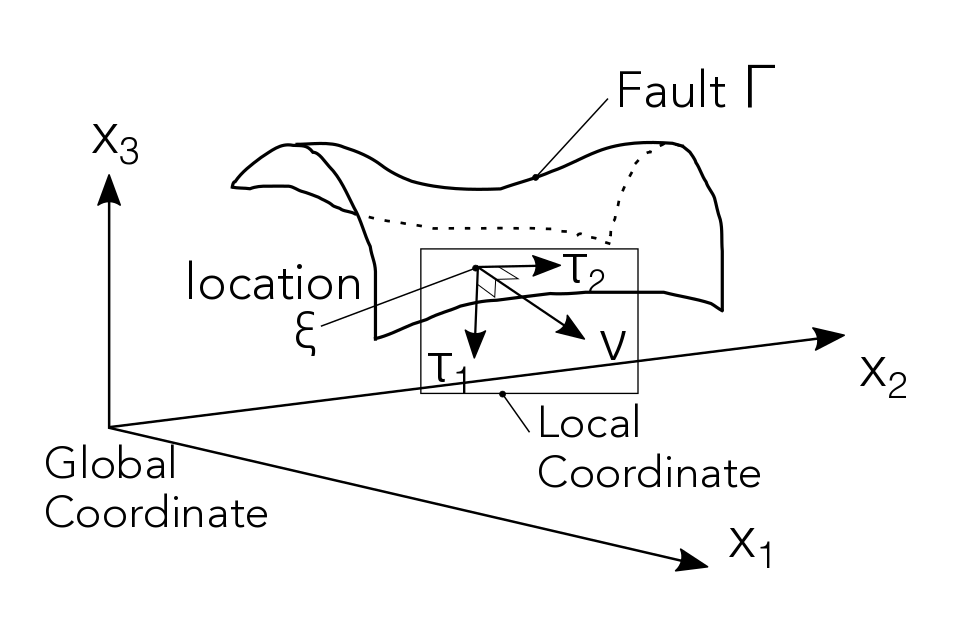}
 \caption{
%(Left panel) 
Schematic illustration of the local coordinate system on a fault $\Gamma$ buried in the global coordinate system with axes $(x_1,x_2,x_3)$. 
The local coordinate system is spanned 
by the normal vector $\boldsymbol\nu$ and the tangential vectors $\boldsymbol\tau_1$ and $\boldsymbol\tau_2$ 
at each location $\boldsymbol\xi$ on $\Gamma$.
 }
\label{fig:schemlocalcoordinate}
\end{figure}

The local coordinate system is a curvilinear, orthonormal coordinate system at the original time, one axis of which is defined by the unit normal vector $\boldsymbol \nu(\boldsymbol\xi,s)$ at the location $\boldsymbol\xi$ on the boundary $\Gamma$ at time $s$.
The other spatial vectors spanning the local coordinate system are unit vectors tangential to the boundary $\Gamma$, denoted by $\boldsymbol\tau_1(\boldsymbol \xi,s), \boldsymbol\tau_2(\boldsymbol\xi,s)$, at the location $\boldsymbol\xi$ and time $s$. 
The tangential vectors $\boldsymbol\tau_1$ and $\boldsymbol\tau_2$ are also orthonormal vectors, which span the tangential plane at each location on $\Gamma$. %, as noticed from that $\boldsymbol\tau_1,\boldsymbol\tau_2$ are orthogonal to the normal vectors $\boldsymbol\nu$ of the fault $\Gamma$.
This local coordinate system is shown schematically in Fig.~\ref{fig:schemlocalcoordinate}.

%断層の各点において、面の法線ベクトルをある軸に取る正規直交座標系(局所座標系)を定める。我々は以外の局所座標系を貼るベクトルを$\tau_1, \tau_2$とかく。$\nu$と$\tau_1,\tau_2$が直交することから、$\tau_1,\tau_2$は断層面を貼るベクトルになっている。

The tangential vectors 
$\boldsymbol\tau_1$ and $\boldsymbol\tau_2$ are arbitrary given at a location on $\Gamma$ as long as they span the plane perpendicular to $\boldsymbol\nu$ as $\boldsymbol\tau_1\times \boldsymbol\tau_2=\boldsymbol\nu$, where ${\bf A}\times{\bf B}$ represents the cross product for given vectors ${\bf A}$ and ${\bf B}$.
%の決定条件は任意である。
For example, 
%たとえば、$x_3$面が解析面に直交する２次元問題では、と選ばれている。
they can be chosen as
$ %\begin{eqnarray}
\boldsymbol
\tau_1
= \hat x_3 \times\boldsymbol\nu,
%\frac{\boldsymbol \nu - (\boldsymbol \nu \cdot \hat x_2)\hat x_2}{|\boldsymbol \nu - (\boldsymbol \nu \cdot \hat x_2)\hat x_2|}, 
\,%\hspace{20pt}
\boldsymbol
\tau_2=\hat x_3 
$ %\end{eqnarray}
in a two-dimensional problem on a plane spanned by the global coordinate axes $x_1$ and $x_2$, 
where $\hat x_a$ represents the unit vector along the $x_a$ axis $(a=1,2,3)$ in the global coordinate system; we can also adopt $\boldsymbol \tau_1= \hat x_2 \times\boldsymbol\nu$ and $\boldsymbol\tau_2=\hat x_2$ for two-dimensional problems on $x_3$-$x_1$ planes, and only $\boldsymbol\tau_1$ is the useful tangential vector for the two-dimensional problem.
The tangential axes
%After $\boldsymbol\tau_1$ and $\boldsymbol\tau_2$ are given at a location on a boundary, 
$\boldsymbol\tau_1$ and $\boldsymbol\tau_2$ at the other locations on the boundary are unambiguously determined through the differential forms given by the geometry of the boundary, as detailed below in \S\ref{sec:323}.
%\footnote{
%Related to the line just before, the following line deleted:
%For example, 
%$\boldsymbol\tau_1$ can be set as the direction of the displacement motion on the boundary, and $\boldsymbol\tau_2$ can be set at the vertical direction of $\boldsymbol\tau_1$. In this case, the local coordinate system depends on time following the time evolution of the displacement.
%%局所座標系は時間変化していても良い。たとえば$\tau_1$を滑りの方向、$\tau_2$を滑りに直行する方向と定めることもできる。この場合滑り方向の時間変化に伴って局所座標系は時間変化している。
%}

%It is useful for the derivation to use the projection of the first, second, foruth order tensors onto the local coordinate.
%導出で我々は、ここまでLatin alphabetsで表してきた1,2,4階のテンソルの添え字に対して、
%局所座標系へ射影した成分表記を用いる。
%なる形で、実座標をなすベクトル との内積denoted by $\cdot $によって定められるi番要素$A_i$のように、  局所座標への射影は と定められる。で表される局所座標系の表記には、我々はアインシュタイン規約はもちいない。
Throughout the following derivation, 
we denote the components of a vector ${\bf A}$ 
along the local coordinate axes by 
\begin{eqnarray}
A_{\nu}&:=&\boldsymbol \nu (\boldsymbol\xi,s)\cdot {\bf A} 
\\
A_{\tau\phi}&:=&\boldsymbol \tau_\phi (\boldsymbol\xi,s) \cdot {\bf A} 
\label{eq:prodtauphi}
\end{eqnarray}
where $\phi$ takes the values 1 or 2. 
It parallels to the definition ($A_i:={\bf A} \cdot \hat x_i$) 
of the $i$-th component $A_i$ in the global coordinate system. 
Higher-order tensors are projected into the local coordinate system in the same way.

While we employ the Latin subscripts for the global coordinate system, we use the Greek alphabet to represent the subscripts in the local coordinate system. 
We do not use Einstein's summation convention for the Greek subscripts. 

Hereafter, we omit the location and time dependences of the local coordinate axes unless the necessity arises.

\subsubsection{Coordinate Values in the Local Coordinate System}
\label{sec:322}
We introduce the coordinate values in the local coordinate system in a differential manner. The differentials are defined simultaneously in the local coordinate. %They play essential roles in the regularization.
%As the real coordinate system, the local coordinate system introduced in \S\ref{sec:321} gives coordinate values to respective locations $\boldsymbol\xi$ on a boundary plane. 
%The transform between those coordinate values in the local and real coordinate systems are written in a differential form, as shown below. The differential form of the coordinate transform consequently gives the differentiation and integration on the local coordinate that play essential roles in the regularization done in \S\ref{sec:33}.

We distinguish a vector parametrized in terms of the local coordinate values, denoted by $\boldsymbol\eta$, from the same vector parametrized in terms of the global coordinate values $\boldsymbol\xi$ to simplify the explanation, as in \citet{tada1997non}. The two vectors $\boldsymbol\eta$ and $\boldsymbol\xi$ represent the same location and are the same vector as long as they are on the boundary where $\boldsymbol\eta$ is defined. 
%Note that the real coordinate value ($\boldsymbol\xi$ in this paper) on $\Gamma$ is denoted by $\boldsymbol\xi$ in \citet{aki2002quantitative} and $\boldsymbol y$ in \citet{tada1996paradox}; the local coordinate value ($\boldsymbol\eta$ in this paper) is denoted by $s$ in \citet{tada1996paradox} and $\boldsymbol \xi$ in \citet{tada1997non}.

%面に沿って導入された局所座標系は、実座標での座標値と異なる位置の座標値(coordinate value)を与える。のちに示すように、これら２つの座標値の微分量を対応づける関係式が我々の結果において本質的に重要となる。ここで、我々は、その関係式を導入する。

%ここでは説明の簡単のため、境界$S$は連続的に繋がっている(SS上のある点は面上の移動によって任意のS上の点に移れる)とする。なお、これが満たされない場合にも、以下の局所座標系の説明\S\ref{sec:32}で微分で表される関係式は成り立つ。
%正則化過程では、断層曲面に沿った部分積分が行われる（T ＆Y,1997）。この面に沿った部分積分は今回の導出の再検討で重要となるため、以下でどのような性質に基づいているかを詳述する。

%局所座標上で、境界面上の位置$\xi$は, ある座標値として指定しうる。
%The convenience to introduce such discrimination becomes clear when they are compared with each other.
%実断層での座標値と局所断層上での座標値は断層面上では等価であるが、T\&Yにならい、我々はこれを実座標値($xi$)とは区別し、ここでは$\eta$とかく。区別の必要性はのちに述べる。

Let the $\phi$ component $\eta_\phi$ of $\boldsymbol\eta$ be the coordinate value along the $\phi$ direction in the local coordinate system. 
Since the $\phi$ direction in the local coordinate system is parallel to $\boldsymbol\tau_\phi$ at $\boldsymbol \xi$, an infinitesimal change $d\eta_\phi$ of $\eta_\phi$ has the following relation to the change $d\boldsymbol\xi$ of $\boldsymbol\xi$ on $\Gamma$,
%$\eta_\phi$は、断層にそって$\phi$の方向に移動した距離として表される。つまり、なる関係を$\xi$との間に持つ。この式によって面上の座標値の変化と実座標上での座標値の変化とが対応づけられる。
\begin{eqnarray}
d\eta_\phi= d\boldsymbol\xi \cdot \boldsymbol\tau_\phi.
\label{eq:linecompetaphi}
\end{eqnarray} 
Eq.~(\ref{eq:linecompetaphi}) connects the coordinate values in the global and local coordinate systems in a differential manner.

Eq.~(\ref{eq:linecompetaphi}) also provides the conversion of the differentiation between the global and local coordinate systems:
\begin{eqnarray}
\partial_{\tau_\phi}
&=& \frac {\partial}{\partial\xi_{\tau_\phi}}= \frac {\partial}{\partial(\boldsymbol\xi\cdot\boldsymbol\tau_\phi)}
\\&=&
\frac{\partial}{\partial\eta_\phi},
\end{eqnarray} 
that is,
\begin{eqnarray}
\partial_{\tau_\phi}
=
\frac{\partial}{\partial\eta_\phi}
\label{eq:convdiff}
\end{eqnarray} 
where we use Eq.~(\ref{eq:prodtauphi}) in the first line 
and Eq.~(\ref{eq:linecompetaphi}) in the transform from the first to the second line. 

%Eq.~(\ref{eq:linecompetaphi}) further relates $d\eta_1,d\eta_2$ to
%an infinitesimal integration area
%$d\Sigma(\boldsymbol\xi)$ in the area integral $\int_\Gamma d\Sigma(\boldsymbol\xi)$ for the real space.
%Since $d\Sigma(\boldsymbol\xi)$ perpendicular to $\boldsymbol\nu$
% is spanned by a small vectors parallel to $\boldsymbol\tau_1$ and $\boldsymbol\tau_2$, given Eq.~(\ref{eq:linecompetaphi}), $d\Sigma(\boldsymbol\xi)$ can be expressed as
%%加えて、$d\Sigma(\boldsymbol\xi)$は$\nu$に直交する $\tau_1,\tau_2$で貼られる面上の微小な面積要素であるため、$d\eta_1,d\eta_2,d\Sigma$の間には なる関係がある。
%\begin{eqnarray}
%d\Sigma(\boldsymbol\xi)
%&=& 
%d\eta_1d\eta_2.
%\label{eq:areacompetaphi}
%\end{eqnarray} 

%As above, the local coordinate is related to the real coordinate. 
%In an opposite way, 
As 
$\boldsymbol\eta$ is parametrized as a function of $\boldsymbol\xi$ on $\Gamma$ by the path integral of Eq.~(\ref{eq:linecompetaphi}), 
$\boldsymbol\xi$ on $\Gamma$ is parametrized as a function of $\boldsymbol\eta$ by the path integral of the following local coordinate expression of $d\boldsymbol\xi$:
%$d\xi(\xi)$は$\nu$に直交することから、$d\xi$の局所座標での成分表示より、
\begin{eqnarray}
d\boldsymbol\xi&=& \boldsymbol\nu (\boldsymbol\nu\cdot d\boldsymbol\xi)+\sum_\phi \boldsymbol\tau_\phi (\boldsymbol\tau_\phi \cdot d\boldsymbol\xi)
\\&=&
\sum_\phi \boldsymbol\tau_\phi d\eta_\phi.
\label{eq:linecompphieta}
\end{eqnarray} 
Here we have used 
Eq.~(\ref{eq:linecompetaphi})
and 
the property 
[$\boldsymbol\nu(\boldsymbol\xi)\cdot d\boldsymbol\xi(\boldsymbol\xi)=0$]
of $d\boldsymbol\xi$ on $\Gamma$--that $d\boldsymbol\xi$ is perpendicular to $\boldsymbol\nu$--for the transform from the first to the second line.
%なる関係もある。ここで第二式で$\nu(\xi)\cdot d\xi(\xi)=0$および式 (直前) を用いた。そこで、$\eta$が$\xi$を式..の両辺の積分によって..とパラメトライズするように、この式の積分によって$\eta$は$\xi$を....とパラメトライズしてもいる。
We utilize this parametrization of $\boldsymbol\xi$ in terms of $\boldsymbol\eta$ in the later numerical experiments. 
%Note that although the surjectivity of the coordinate transform from the local coordinate to the real coordinate is not obvious in three-dimensional problems for randomly chosen sets of $\boldsymbol\tau_\phi$ directions, such surjectivity is not significant for constructing the non-hypsersingular expressions of displacement gradient and stress Green's functions.
%%Eqs.~(\ref{eq:linecompetaphi}) and (\ref{eq:linecompphieta}) are path dependent 
%%$\eta$による$\xi$のパラメトライズは...で用いられる。%ただし、２式..x.. の積分値は、2次元面では経路依存量として求められる。そこで、3次元媒質(3次元問題)での場合では、局所座標系が時間不変の場合でさえ、$\eta$による$\xi$、あるいは$\xi$による$\eta$のパラメトライズはそれほど単純ではない。2次元媒質(2次元問題)での場合は、あとで図..に示されるように、比較的簡易的に解を求めることができる。
%While the coordinate value $\boldsymbol\eta$ in the local coordinate is not defined to the locations off the boundary, Eqs.~(\ref{eq:linecompetaphi}) and (\ref{eq:linecompphieta}) allow us to identify 
%the vectors $\boldsymbol\eta$ and $\boldsymbol\xi$ pointing a same location
%as long as $\boldsymbol\xi$ are located on the boundary.
%断層面上での位置ベクトルを異なる２つの座標系で示したものに過ぎない。したがってこれらの量は等価である。しかし、面外の座標値$x$は局所座標では定義されないため、グリーン関数の中で断層面外での位置座標との差を取る演算を局所座標は扱えない。このような定義域の違いを踏まえると、その際に$\eta$と$\xi$の区別を行なっておくと見通しが良い。なお、実座標での断層座標(我々の表記での$\xi$)は、ARでは$\xi$、TYでは$y$と表されている。局所座標(我々の表記ではeta)はTY1996では$s$,TY1997では$xi$と表されている。

\subsubsection{
Stokes' Theorem 
and 
Integration-by-Parts Technique 
%Parametrization of Coordinate Axes as Functions of Coordinate Values
}
\label{sec:323}
Last, we introduce an integral formula that holds in arbitrary local coordinates in order to regularize the stress Green's function. 

Stokes' theorem connects the boundary integral of $\nabla \times {\bf F}$ on $\Gamma$ with the line integral of a vector function ${\bf F}$ tracing the edge $\partial \Gamma$ of $\Gamma$: 
\begin{eqnarray}
\int_{\Gamma} d\Sigma (-\boldsymbol\nu) \cdot (\nabla \times {\bf F})
=
\int_{\partial\Gamma} d\boldsymbol\Lambda \cdot {\bf F},
\label{eq:stokes1}
\end{eqnarray}
where $d\boldsymbol\Lambda$ denotes an infinitesimal vector parallel to the direction of motion along the path of integration. 
Eq.~(\ref{eq:stokes1}) can be rewritten in the tensorial form 
\begin{eqnarray}
-\int_{\Gamma} d\Sigma \nu_a \epsilon_{abc} \partial_b F_c
=
\int_{\partial\Gamma} d\Lambda_a F_a,
\label{eq:stokes2}
\end{eqnarray}
where $\epsilon_{abc}$ denotes the Levi-Civita symbol, which yields 
$\epsilon_{abc}=1 $ when $(a,b,c)=(1,2,3)$, $(2,3,1)$, or $(3,1,2)$; 
$\epsilon_{abc}=-1 $ when $(a,b,c)=(3,2,1)$, $(1,3,2)$, or $(2,1,3)$; 
and 
$\epsilon_{abc}=0$ otherwise. 

When ${\bf F}$ is given by a scalar function $f$ as $F_a=\delta_{ad}f$, Eq.~(\ref{eq:stokes2}) becomes
\begin{eqnarray}
- \int_{\Gamma} d\Sigma \epsilon_{abd} \nu_a  \partial_b f
=
\int_{\partial\Gamma} d\Lambda_d f
\label{eq:stokes3}
\end{eqnarray}
Further multiplying Eq.~(\ref{eq:stokes3}) by $\epsilon_{jmd}$ and using 
\begin{eqnarray}
\epsilon_{jmd}\epsilon _{abd}=\delta_{ja}\delta_{mb}-
\delta_{jb}\delta_{ma},
\end{eqnarray}
we get the following: 
\begin{eqnarray}
-\int_{\Gamma} d\Sigma 
(\nu_j\partial_m-\nu_m\partial_j)f
=
\int_{\partial\Gamma} d\Lambda_d \epsilon_{jmd}f.
\label{eq:stokes4}
\end{eqnarray}
Due to its antisymmetric property 
[$(\nu_j\partial_m-\nu_m\partial_j)= -(\nu_m\partial_j-\nu_j\partial_m)$], 
the tensor $\nu_j\partial_m-\nu_m\partial_j$ in Eq.~(\ref{eq:stokes4}) is expressed by the differentials in the local coordinates as 
\begin{eqnarray}
&&\nu_j\partial_m-\nu_m\partial_j 
\nonumber\\&=&
\nu_j(\partial_m -\nu_m\partial_\nu+ \nu_m\partial_\nu)
%\nonumber\\&&
-\nu_m(\partial_j-\nu_j\partial_\nu+ \nu_j\partial_\nu)
\\&=&
\nu_j(\partial_m -\nu_m\partial_\nu) -\nu_m(\partial_j-\nu_j\partial_\nu)
\label{eq:stokes4to5}
\end{eqnarray}
We then utilize a property of the inner product; for two given vectors ${\bf A}$ and ${\bf B}$, the inner product satisfies the following relation: 
%以降の変換は内積の座標不変性を用いることで見通しよく実行できる。ベクトルA,B に対して、内積は以下の性質を持つ（Appendix?）。%なお、これらの添え字はいずれも$x$に依存しないため、 A,Bの中に微分演算子が含まれていても成り立つことがわかる。
\begin{eqnarray}
{\bf A}\cdot {\bf B}=A_m B_m=A_\nu B_\nu +\sum _\phi A_{\tau_\phi}B_{\tau_\phi}.
\label{eq:inproindep}
\end{eqnarray}
By substituting ${\bf A} =\hat x_m$ and ${\bf B}=\boldsymbol\nabla$ to Eq.~(\ref{eq:inproindep}) for a unit vector $\hat x_m$ in the $m$-th direction in the global coordinate system, 
we obtain component expressions of the differential in the global and local coordinates:   
%式xxの左辺に含まれる微分$\partial_m$が、$\partial_\nu$となる場合、式xx はxxでのそこで、$\partial_m$を局所座標で表現することで$\partial_\nu$を取り出す。式xxに、$A =\hat x_m$, $B=\boldsymbol\nabla$を代入するとを得る。
%ここで$\tau_{\phi m}$は$\tau_{\phi}$の第m成分. ここで得られた、$\partial_m$の局所座標系での表現を式xxに代入すると、
\begin{eqnarray}
\hat x_m \cdot \boldsymbol\nabla=\partial_m=\nu_m\partial_\nu+\sum_\phi \tau_{\phi m} \partial_{\phi},
\label{eq:projnabla}
\end{eqnarray}
where $\tau_{\phi m}$ is the $m$-th component of $\tau_{\phi}$.
With this component expression of the differential in the global and local coordinates, successive calculations to obtain Eq.~(\ref{eq:stokes4to5}) lead to 
\begin{eqnarray}
\nu_j\partial_m-\nu_m\partial_j =
\sum_{\phi} (\nu_j\tau_{\phi m} - \nu_m\tau_{\phi j})\partial_{\tau_\phi}
\label{eq:stokes5}
\end{eqnarray}
Eqs.~(\ref{eq:stokes4}) and (\ref{eq:stokes5}) give the following relation for the linear operator $\sum_{\phi} (\nu_j\tau_{\phi m} - \nu_m\tau_{\phi j})\partial_{\tau_\phi}$:
\begin{eqnarray}
\sum_{\phi} \int_{\Gamma} d\Sigma 
(\nu_j\tau_{\phi m} - \nu_m\tau_{\phi j})\partial_{\tau_\phi} f
=
-\int_{\partial\Gamma} d\Lambda_d 
\epsilon_{jmd}
f.
\label{eq:stokes6}
\end{eqnarray}

Let $f$ equal to the product $gh$ of two given functions $g$ and $h$. 
As long as $gh=0$ at the edges $\partial \Gamma$ of the boundary $\Gamma$, 
the path integral in Eq.~(\ref{eq:stokes6}) vanishes as
\begin{eqnarray}
\sum_{\phi} \int_{\Gamma} d\Sigma 
(\nu_j\tau_{\phi m} - \nu_m\tau_{\phi j})\partial_{\tau_\phi} (gh)
=
0,
\end{eqnarray}
or equivalently,
\begin{eqnarray}
&&\sum_{\phi} \int_{\Gamma} d\Sigma 
g(\nu_j\tau_{\phi m} - \nu_m\tau_{\phi j})\partial_{\tau_\phi} h
\nonumber\\&&=
-\sum_{\phi} \int_{\Gamma} d\Sigma 
h(\nu_j\tau_{\phi m} - \nu_m\tau_{\phi j})\partial_{\tau_\phi} g.
\label{eq:stokes7}
\end{eqnarray}
Note that the edge condition $gh=0$ can be interchanged with the condition of continuity of $gh$ if the boundary is periodic. 
Eq.~(\ref{eq:stokes7}) can be regarded as a kind of integration by parts and is used in the derivation [this is the so-called ``integration-by-parts technique''~\citep[p21]{bonnet1999boundary}], although it is not a naive integration by parts obtained from the divergence theorem. 
Eq.~(\ref{eq:stokes7}) holds on the respective unjointed boundaries even when $\Gamma$ is made of multiple unjointed boundaries. 

\begin{figure}%[tbp]
  \includegraphics[width=75mm]{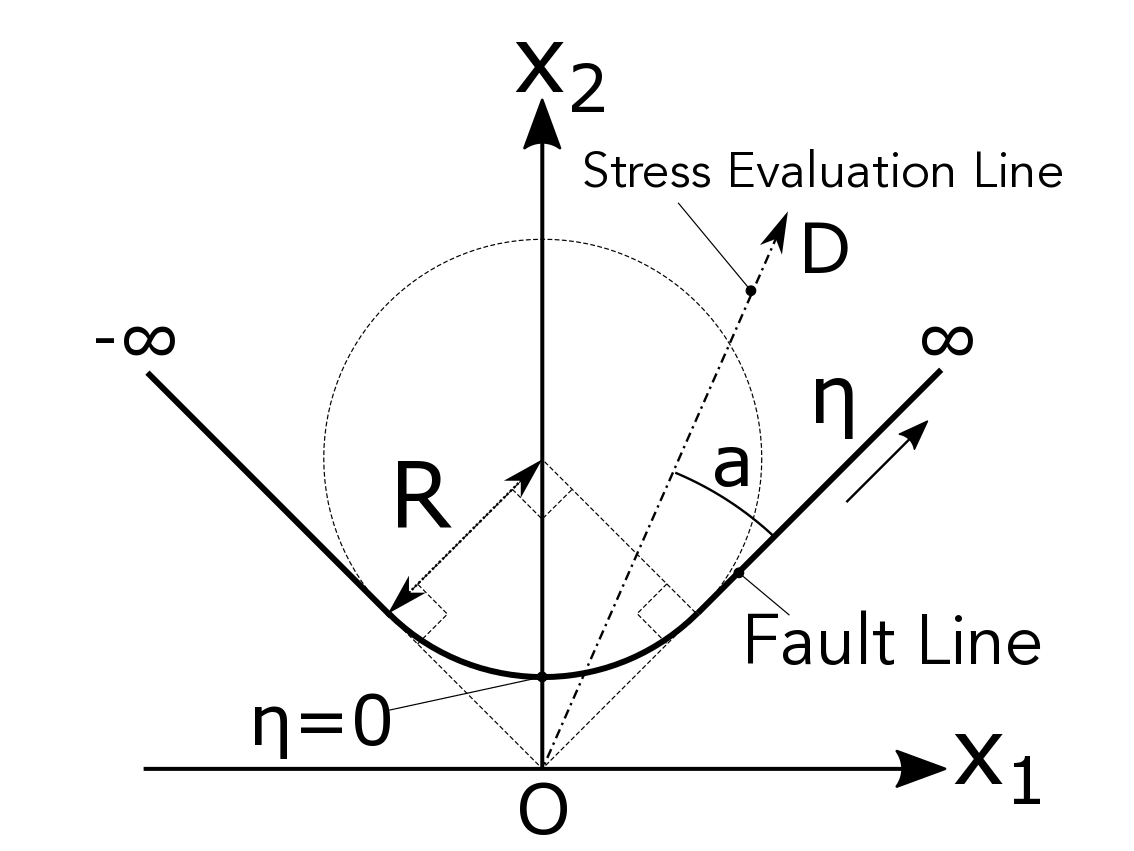}
 \caption{
%(Left panel) 
Simulated fault geometry. Two half-lines connect to the edges of the arc of a circle of radius $R$, subtending an angle $\pi/2$, where the half-lines coincide with the tangents to the arc. The origin of the global coordinate system is located at the intersection of the extensions of the half-lines. The fault is symmetric about the $x_2$ axis. 
%The origin of the local coordinate is set at the intersection of the fault line and $x_2$ axis. 
The off-fault stress is evaluated in the numerical experiments along the stress evaluation line shown, which is parametrized by an angle $a$. %The value of $a$ is set at $a=\mbox{atan}(2)-\pi/4$.
%(Right panel) schmatic of junction of piecewise constant (when we take a limit $\delta x/\Delta x\to 0$).
 }
\label{fig:1}
\end{figure}

As shown in Eq.~(\ref{eq:stokes7}), the two tensors $(\nu_j\tau_{\phi m} - \nu_m\tau_{\phi j})$ and $\partial_{\tau_\phi}$ is permutable in the boundary integral over $\Gamma$, given the condition $gh=0$ on $\partial\Gamma$. This is nontrivial, given that the local coordinate axes evolve in space. 
For example, 
let $\Gamma$ be the circular arc in Fig.~\ref{fig:1}, which has radius $R$ and forms an angle $\pi/2$ in two-dimensional $x_1-x_2$ space.
The normal vector $\boldsymbol\nu$ is location dependent, and it changes smoothly while retaining the orthonormal relation to the tangential vector $\boldsymbol\tau=\boldsymbol\tau_1$ as the local coordinate value $\eta=\eta_1$ evolves. The axes thus evolve along the curve in a differential manner:
\begin{eqnarray}
\partial_{\tau_\phi} \boldsymbol\tau_\phi &=& 
\frac {d\boldsymbol\tau }{d\eta}=R^{-1} \boldsymbol\nu 
\label{eq:curvatureproptau}
\\
\partial_{\tau_\phi} \boldsymbol\nu &=&
\frac {d\boldsymbol\nu }{d\eta}=-R^{-1} \boldsymbol\tau
\label{eq:curvaturepropnu}
\end{eqnarray} 
where we have used Eq.~(\ref{eq:convdiff}) to express the differentials of the local coordinate axes in both the global and local coordinates. 
%These describe the evolution of the local coordinate axes depending on the geometry of the arc. Indeed, t
Integration of Eqs.~(\ref{eq:curvatureproptau}) and (\ref{eq:curvaturepropnu}) gives the differences in the local coordinate axes at different places along the arc.
As above, the differentials of $\boldsymbol \tau_\phi$ and $\boldsymbol\nu$ are exactly non-zero, despite the fact that $(\nu_j\tau_{\phi m} - \nu_m\tau_{\phi j})$ is substantially permutable with the differential operator $\partial_{\tau_\phi} $ in Eq.~ (\ref{eq:stokes7}) due to Stokes' theorem, given the edge condition on $gh$. Eq.~ (\ref{eq:stokes7}) holds for problems in any dimension, although $\sum_\phi \partial_{\tau_\phi} (\nu_j\tau_{\phi m} - \nu_m\tau_{\phi j})$ can take locally non-zero values in three-dimensional cases.

\subsection{Regularization of the Displacement-Gradient and Stress Green's Functions}
 \label{sec:33}
We here regularize the integral equation, Eq.~(\ref{eq:displacementgradienthom}), that describes the displacement-gradient field, eventually giving the hypersingular expression of the stress Green's function on the fault $\Gamma$. 
%The regularization of $D_{4,mn}$ eventually gives the non-hypersingular expressions of the displacement gradient and stress Green's function. 
To begin with, we simplify Eq.~(\ref{eq:displacementgradienthom}) by using the spatial translational symmetry [Eq.~(\ref{eq:spatialtrans})] of the Green's function ${\bf G}$ in a homogeneous medium:
\begin{eqnarray}
\partial_m u_{n}({\bf x},t) 
&=& -
\int^{\infty}_{-\infty} ds 
\int_{\Gamma(s)} d\Sigma(\boldsymbol\xi) 
\Delta u_{i}(\xi,s) 
\nonumber\\&&\times
\nu_j(\boldsymbol\xi,s) c_{ijpq} 
\partial^{(\xi)}_m \partial^{(\xi)}_q G_{np} ({\bf x},t;\boldsymbol\xi,s).
\label{eq:D4hom}
\end{eqnarray}
%Structuring the regularization techniques in the previous studies (e.g., Bonnet,xx; Koller et al., 199x), we show a general way of regularization with using the equation of motion and the partial integration along the boundary.
%式xxでの$c_{i\nu pq} \partial_m \partial_q G_{np}$は運動方程式と加速度項を抜いた部分と添え字を除いて一致している。そこで以下ではこの項に似た部分を運動方程式から取り出すことを考える。

We suppose the evaluation point ${\bf x}$ to be an off-fault location (${\bf x}\notin\Gamma$) that never coincides with the location of the source $\boldsymbol\xi$ on the fault $\Gamma$ (${\bf x}\neq \boldsymbol\xi\in \Gamma$) in Eq.~(\ref{eq:displacementgradienthom}). This enables us to avoid considering the delta-function contained in the equation of motion Eq.~(\ref{eq:EOMofG}) of the Green's function. 
Even under such an assumption, the stress on the fault can be evaluated by using the continuity of the stress near the fault [$\sigma(\boldsymbol \xi,t)= \sigma(\boldsymbol \xi\pm0\times \boldsymbol\nu, t)$ for ${\bf x}\in \Gamma$]. 
This parallels the handling of Eq.~(\ref{eq:EOMofG}) in previous studies, which did not evaluate the delta-functions in Eq.~(\ref{eq:EOMofG}) when considering the on-fault stress~\citep[e.g.,][]{tada1997non}. %式..では$\xi$が断層面上、$x(\neq \xi)$は断層面外にある(as mentioned in xx)。そこで 式..Green関数の運動方程式xxからデルタ関数が消去される。面直上の評価にはデルタ関数の評価が要求されるため、ここでは扱われない。

For a source and receiver in different locations, the equation of motion, Eq.~(\ref{eq:EOMofG}), of the Green's function in a homogeneous medium ($\partial_jc_{ijpq}=0$) reduces to 
%これをグリーン関数の従う運動方程式
\begin{eqnarray}
\rho \partial_t^2 G_{in}({\bf x},t;\boldsymbol \xi,s)  %&=&c_{ijqp}\partial_j\partial_q G_{np}({\bf x},t) \label{eq:EOMGofffbefore}\\
&=&c_{ijpq}\partial_j^{(x)} \partial_q^{(x)} G_{pn}({\bf x},t;\boldsymbol \xi,s),
\label{eq:EOMGofff}
\end{eqnarray}
where we have used ${\bf x}\neq \boldsymbol\xi$, the spatial homogeneity of the elasticity tensor $c$, and a symmetry of $c$: $c_{ijpq}=c_{ijqp}$.
%ここで第二行目への変換で$c_{ijpq}=c_{ijqp}$なる対称性と$c$の空間一様性を用いた。式xxの右辺について、式xxで、$A_i=c_{ijpq}$, $B_i=\partial_j \partial_q G_{np}({\bf x},t)$とすると、

\subsubsection{Non-Hypersingular Displacement-Gradient Green's Functions}
We obtain the regularized form of the displacement gradient below.
Projecting the $j$-th component in Eq.~(\ref{eq:displacementgradienthom}) into the local coordinate system, we obtain 
\begin{eqnarray}
\partial_mu_{n}({\bf x},t) 
&=& -
\int^\infty_{-\infty} ds \int_{\Gamma(s)} d\Sigma(\boldsymbol\xi) 
\Delta u_{i}(\boldsymbol\xi,s) 
\nonumber\\&&\times
c_{i\nu pq} 
\partial^{(\xi)}_m \partial^{(\xi)}_q G_{np} ({\bf x},t;\boldsymbol\xi,s).
\label{eq:D4homproj}
\end{eqnarray}
This expression has the subscript $\nu$ of $c_{i\nu pq}$ expressed in local coordinates, while the other subscripts are expressed in global coordinates.
As in this expression, the regularized form of the displacement gradient is expressed with subscripts along the axes of both the global and local coordinate systems. 

The term $c_{i\nu pq} \partial_m \partial_q G_{np}$ in Eq.~(\ref{eq:D4homproj}) contains the $m=\nu$ component $c_{i\nu pq} \partial_\nu \partial_q G_{np}$, which is a part of the stress produced by the Green's function; 
%Then it is noticed that when $\nu=m$, 
%in Eq.~(\ref{eq:D4homproj}) coincides with the $\nu$ component of the stress made by the Green's function, 
this can be explicitly written with the spatial reciprocity, Eq.~(\ref{eq:spatialrecipro}), as 
\begin{eqnarray}
c_{ij pq} 
\partial^{(\xi)}_j \partial^{(\xi)}_q G_{np} ({\bf x},t;\boldsymbol\xi,s)
=
c_{ij pq} 
\partial^{(\xi)}_j \partial^{(\xi)}_q G_{pn} (\boldsymbol\xi,t;{\bf x},s).
\label{eq:stressrecipro}
\end{eqnarray}
We here flipped the subscripts $n$ and $p$ and the locations ${\bf x}$ and $\boldsymbol \xi$ in Eq.~(\ref{eq:spatialrecipro}) for comparing Eq.~(\ref{eq:stressrecipro}) with Eq.~(\ref{eq:D4homproj}).
The left-hand side of Eq.~(\ref{eq:stressrecipro}) is certainly $c_{i\nu pq} \partial_\nu \partial_q G_{np}$ in Eq.~(\ref{eq:D4homproj}) for $j=\nu$, and is also a part of the stress produced by the Green's function, as expressed in the right-hand side of Eq.~(\ref{eq:stressrecipro}).

To distinguish the case $\nu=m$ explicitly from the other cases of $c_{i\nu pq} \partial_m \partial_q G_{np}$, we project the subscript $m$ in the partial-differential operator $\partial_m$ in Eq.~(\ref{eq:D4homproj}) into local coordinates with Eq.~(\ref{eq:projnabla});
\begin{eqnarray}
\partial_m u_{n}({\bf x},t) 
&&
= -
\int^\infty_{-\infty} ds\int_{\Gamma(s)} d\Sigma(\boldsymbol\xi)
\Delta u_{i}(\boldsymbol\xi,s) c_{i\nu pq} 
\nonumber\\&&
(\nu_m\partial^{(\xi)}_\nu +\sum_\phi \tau_{\phi m}\partial^{(\xi)}_{\tau\phi} )
\partial^{(\xi)}_q G_{np} ({\bf x},t;\boldsymbol\xi,s).
%\\&=&
%-
%\int^\infty_{-\infty} ds\int_{\Gamma(s)} d\Sigma(\boldsymbol\xi)
%\Delta u_{i}(\boldsymbol\xi,s)  \nu_m [c_{i\nu pq}\partial^{(\xi)}_\nu \partial^{(\xi)}_q G_{np} ({\bf x},t;\boldsymbol\xi,s)]
%\nonumber\\&&
%-\sum_\phi 
%\int^\infty_{-\infty} ds
%\int_{\Gamma(s)} d\Sigma(\boldsymbol\xi) 
%\Delta u_{i}(\boldsymbol\xi,s) \tau_{\phi m}[c_{i\nu pq}\partial^{(\xi)}_{\tau\phi} \partial^{(\xi)}_q G_{np} ({\bf x},t;\boldsymbol\xi,s)].
\label{eq:D4homproj2}
\end{eqnarray}
The first term corresponds to $c_{i\nu pq} \partial_\nu \partial_q G_{np}$ mentioned earlier. The second term is able to be integrated by part. Then intractable hypersingularity is noticed to be in the first term. 
 
The part proportional to $\partial_\nu$ in Eq.~(\ref{eq:D4homproj2}) is transformed by the equation of motion, Eq.~(\ref{eq:EOMGofff}), of the Green's function.
The transformation takes the following four steps.
First, Eq.~(\ref{eq:stressrecipro}) and 
Eq.~(\ref{eq:EOMGofff}) of flipping ${\bf x}$ and $\boldsymbol\xi$ 
give 
\begin{eqnarray}
c_{ij pq} 
\partial^{(\xi)}_j \partial^{(\xi)}_q G_{np} ({\bf x},t;\boldsymbol\xi,s)
=
\rho \partial_t^2 G_{in}(\boldsymbol\xi,t;{\bf x},s).
\label{eq:EOMGofffchanged}
\end{eqnarray}
Second, substituting $A_j=c_{ijpq}$ and $B_j=\partial^{(\xi)}_j \partial^{(\xi)}_q G_{np}({\bf x},t;\boldsymbol\xi,s)$ into Eq.~(\ref{eq:inproindep}) of replacing $m$ with $j$, 
we can separate $c_{ijpq}\partial_j\partial_p G_{nq}$ into  
\begin{eqnarray}
&&c_{ijpq}\partial^{(\xi)}_j \partial^{(\xi)}_q G_{np}({\bf x},t;\boldsymbol\xi,s)
\nonumber\\
&=& 
c_{i\nu pq}\partial^{(\xi)}_\nu \partial^{(\xi)}_q G_{np}({\bf x},t;\boldsymbol\xi,s)
\nonumber\\&&
+\sum_\phi c_{i\tau_\phi pq}\partial^{(\xi)}_{\tau_\phi} \partial^{(\xi)}_q G_{np}({\bf x},t;\boldsymbol\xi,s). 
\label{eq:projsigG}
\end{eqnarray}
Third, using Eqs.~(\ref{eq:EOMGofffchanged}) and (\ref{eq:projsigG}), 
%we continue the calculation expressed by Eqs.~(\ref{eq:EOMGofffbefore}) 
we find
%\begin{eqnarray}
%\rho \partial_t^2 G_{ni}({\bf x},t;\boldsymbol\xi,s)= 
%c_{i\nu pq}\partial_j \partial_\nu G_{np}({\bf x},t;\boldsymbol\xi,s)+
%\sum_\phi c_{i\tau_\phi pq}\partial_{\tau_\phi} \partial_q G_{np}({\bf x},t;\boldsymbol\xi,s), 
%\end{eqnarray}
%or equivalently, 
\begin{eqnarray}
&&c_{i\nu pq}\partial^{(\xi)}_\nu \partial^{(\xi)}_q G_{np}({\bf x},t;\boldsymbol\xi,s)=
\nonumber\\&&
\rho \partial_t^2 G_{in}(\boldsymbol\xi,t;{\bf x},s)
-\sum_\phi c_{i\tau_\phi pq}\partial^{(\xi)}_{\tau_\phi} \partial^{(\xi)}_q G_{np}({\bf x},t;\boldsymbol\xi,s). 
\label{eq:projEOMGofff0}
\end{eqnarray}
Fourth, we rewrite Eq.~(\ref{eq:projEOMGofff0}) with using the temporal translational symmetry, Eq.~(\ref{eq:temporaltrans}), and the spatial reciprocity, Eq.~(\ref{eq:spatialrecipro}), of the Green's function in the form 
\begin{eqnarray}
&&c_{i\nu pq}\partial^{(\xi)}_\nu \partial^{(\xi)}_q G_{np}({\bf x},t;\boldsymbol\xi,s)=
\nonumber\\&&
\rho \partial_s^2 G_{ni}({\bf x},t;\boldsymbol\xi,s)
-\sum_\phi c_{i\tau_\phi pq}\partial^{(\xi)}_{\tau_\phi} \partial^{(\xi)}_q G_{np}({\bf x},t;\boldsymbol\xi,s). 
\label{eq:projEOMGofff}
\end{eqnarray}
We here utilized the abovementioned assumption that a source and receiver are at different locations.

The first term in Eq.~(\ref{eq:D4homproj2}) is proportional to the left hand side of Eq.~(\ref{eq:projEOMGofff}). 
Substituting Eq.~(\ref{eq:projEOMGofff}) into Eq.~(\ref{eq:D4homproj2}), we get
\begin{eqnarray}
&&
\partial_m u_{n}({\bf x},t) 
\nonumber\\&=&
-\int^\infty_{-\infty} ds \int_{\Gamma(s)} d\Sigma(\boldsymbol\xi)\Delta u_{i}(\boldsymbol\xi,s)
\nonumber\\&&
\left[
\nu_m 
\left\{
\rho \partial_s^2 G_{ni}({\bf x},t;\boldsymbol\xi,s)
\right.
\right.
\nonumber\\&&
\left.
-\sum_\phi c_{i\tau_\phi pq}
\partial^{(\xi)}_{\tau_\phi} \partial^{(\xi)}_q G_{np}({\bf x},t;\boldsymbol\xi,s)
\right\}
\nonumber\\&&
\left.
+\sum_\phi 
c_{i\nu pq} \tau_{\phi m}\partial^{(\xi)}_{\tau\phi} 
\partial^{(\xi)}_q G_{np} ({\bf x},t;\boldsymbol\xi,s).
\right]
\end{eqnarray}
Collecting the terms proportional to 
$\partial^{(\xi)}_{\tau_\phi} \partial^{(\xi)}_q G_{np}({\bf x},t;\boldsymbol\xi,s)$, this is arranged as
%この式の右辺第一項のかっこ[]の部分に式xxを代入すると、
\begin{eqnarray}
&&
\partial_m u_{n}({\bf x},t) 
%-
%\int^\infty_{-\infty} ds\int_{\Gamma(s)} d\Sigma(\boldsymbol\xi) \Delta u_{i}(\boldsymbol\xi,s)\nu_m \rho \partial_s^2 G_{ni}({\bf x},t;\boldsymbol\xi,s)
%\nonumber\\&&
%+\sum_\phi\int^\infty_{-\infty} ds\int_{\Gamma(s)} d\Sigma(\boldsymbol\xi)
%\Delta u_{i}(\boldsymbol\xi,s)  \nu_m [c_{i\tau_\phi pq}\partial^{(\xi)}_{\tau_\phi} \partial^{(\xi)}_q G_{np} ({\bf x},t;\boldsymbol\xi,s)]
%\nonumber\\&&
%-\sum_\phi \int^\infty_{-\infty} ds\int_{\Gamma(s)} d\Sigma(\boldsymbol\xi) 
%\Delta u_{i}(\boldsymbol\xi,s) \tau_{\phi m}[c_{i\nu pq}\partial^{(\xi)}_{\tau_\phi} \partial^{(\xi)}_q G_{np} ({\bf x},t;\boldsymbol\xi,s)].
%\\&=&
\nonumber\\&=&
-\int^\infty_{-\infty} ds \int_{\Gamma(s)} d\Sigma(\boldsymbol\xi)\Delta u_{i}(\boldsymbol\xi,s)\nu_m \rho \partial_s^2 G_{ni}({\bf x},t;\boldsymbol\xi,s)
\nonumber\\&&
-\sum_\phi\int^\infty_{-\infty} ds \int_{\Gamma(s)} d\Sigma(\boldsymbol\xi) 
\Delta u_{i}(\boldsymbol\xi,s)  c_{ij pq}
\nonumber\\&&\times
(\nu_j\tau_{\phi m} -\nu_m\tau_{\phi j})
\partial^{(\xi)}_{\tau\phi} \partial^{(\xi)}_q G_{np} ({\bf x},t;\boldsymbol\xi,s).
\label{eq:D4homproj3}
\end{eqnarray}
We here used $c_{i\nu pq}=c_{ijpq}\nu_j$, $c_{i\tau_\phi pq}=c_{ijpq}\tau_{\phi j}$. 
Since the derivative in Eq.~(\ref{eq:D4homproj3}) contains only the derivative with respect to the time $s$ or the spatial derivative along the boundary $\Gamma$, Eq.~(\ref{eq:D4homproj3}) can be integrated by parts, given Eq.~(\ref{eq:convdiff}): $\partial^{(\xi)}_{\tau\phi}=\partial/(\partial \eta_\phi)$.
Furthermore, 
the operator  
$(\nu_j\tau_{\phi m} -\nu_m\tau_{\phi j})
\partial^{(\xi)}_{\tau\phi}$ in Eq.~(\ref{eq:D4homproj3})
is indeed what is treated by Eq.~(\ref{eq:stokes7}), and we can use the integration-by-parts technique mentioned earlier.  
From these considerations, integrating by parts for the time $s$ in the first term and applying Eq.~(\ref{eq:stokes7}) to %the $\tau_\phi$ component of the space in 
the second term, 
Eq.~(\ref{eq:D4homproj3})
reduces to
%右辺第1項で時間$s$について、第二項で空間$\tau_\phi$方向について部分積分を実行することで、
\begin{eqnarray}
&&\partial_m u_{n}({\bf x},t) 
\nonumber\\&=&
\int^\infty_{-\infty} ds\int_{\Gamma(s)} d\Sigma(\xi)\Delta \dot u_{i}(\boldsymbol\xi,s)\nu_m \rho \partial_s G_{ni}({\bf x},t;\boldsymbol\xi,s)
\nonumber\\&+&
\sum_\phi
\int^\infty_{-\infty} ds\int_{\Gamma(s)} d\Sigma(\boldsymbol\xi)
\frac{\partial
\Delta u_{i}}
{\partial \xi_{\tau_\phi} }
(\boldsymbol\xi,s)  
\nonumber\\&&\times
(\nu_j\tau_{\phi m} -\nu_m\tau_{\phi j})c_{ij pq}\partial^{(\xi)}_q G_{np} ({\bf x},t;\boldsymbol\xi,s),
\label{eq:D4reducedA}
\end{eqnarray}
where the quantity $\Delta \dot u_i(\boldsymbol\xi,s):=\partial_s \Delta u_i(\boldsymbol\xi,s)$ denotes the $i$-th component of the slip rate at location $\boldsymbol\xi$ and time $s$.
In the spatial integration by part, the contribution from the endpoints of the integral vanishes, due to $\Delta {\bf u}={\bf 0}$ at the edge for finite-sized boundaries, ${\bf G}={\bf 0}$ at infinity for infinitely long boundaries, and the continuity of $\Delta {\bf u}$ for periodic boundaries; the temporal one also vanishes due to ${\bf G}={\bf 0}$ in the infinite past and the infinite future. 
Although the partial integration over the time $s$ affects the boundary geometry $\Gamma$ and $\boldsymbol\nu$ both depending on $s$, such effects are expressed by the product of the slip and the temporal rate of change in $\Gamma$, and is the negligible second order under the assumption of small deformations introduced initially.  
%ここで被積分関数が端で0であることを用いた。

Eq.~(\ref{eq:D4reducedA}) is the desired regularized expression for the hypersingular displacement-gradient Green's function, Eq.~(\ref{eq:displacementgradienthom}), which we have obtained after partial integrations along the boundary and in time.
The derivative of the Green's function can be found in \citet{tada2000non} for three-dimensional isotropic media and partly in \citet{tada1997non} for two-dimensional isotropic media.
The term proportional to the time derivative of the slip corresponds to an equivalent single force caused by the inertial effect $\partial_s^2u_i$ in the equation of motion. 
The other term is proportional to the spatial derivative along the boundary and corresponds to the displacement-gradient field caused by the dislocation, which has been studied intensively in the ordinary literature of dislocation theory.
Eq.~(\ref{eq:D4reducedA}) is much more compact than previously obtained expressions that contain dozens of terms~\citep[e.g.,][]{tada2006stress}. 
Please refer to Appendix \ref{sec:comparisonofnomenclature} for differences in notations between this study and other mentioned studies.

Eq.~(\ref{eq:D4reducedA}) is indeed equivalent to the result of \citet{bonnet1999boundary} (p176), as examined in the discussion section. 
%while \citet{bonnet1999boundary} dropped the third term, probably unintentionally. 
%Although the third term is probably a newly found effect missed in the previously obtained stress Green's functions,
%since our first motivation is the paradox of curved faults found even in two-dimensional problems where the third term vanishes, 
%we do not go into the further consideration on the third term of Eq.~(\ref{eq:D4reduced}) in this paper; 
%we will briefly mention the effect of the third term in the discussion section. 
Note that \citet{fukuyama1998rupture} reported that the hypersingularity is caused by the S-wave part in the Green's function, and they regularized the S-wave part only, at least for the shear-dislocation problem. 
Our result is partially integrated over all the other stress fields, which are caused by the P-wave and near-field terms in addition to that caused by the S-wave part for the case of an isotropic medium. This makes our expression applicable to general homogeneous media and not just to an isotropic medium. 
Although our result in Eq.~(\ref{eq:D4reducedA}) may include unnecessary integration by parts in that sense,
the same discretized results will be obtained from Eq.~(\ref{eq:D4reducedA}) for the shear-dislocation problem in an isotropic medium as from the previous studies that regularized the S-wave part only~\citep[e.g.,][]{fukuyama1998rupture,aochi2000spontaneous,tada2006stress}, since both safely handle the hypersingular part of the S-wave.

Eq.~(\ref{eq:D4reducedA}) above can be reduced to a form appropriate for two-dimensional problems by considering a boundary and slip distribution that are translationally symmetric for a given $\tau_2$ direction:
\begin{eqnarray}
&&\partial_m u_{n}({\bf x},t) 
\nonumber\\
&=&
\int^\infty_{-\infty} ds\int_{\Gamma(s)} d\Sigma(\boldsymbol\xi) \Delta \dot u_{i}(\boldsymbol\xi,s) \nu_m\rho \partial_s G_{ni}({\bf x},t;\boldsymbol\xi,s)
\nonumber\\&&
+\int^\infty_{-\infty} ds\int_{\Gamma(s)} d\Sigma(\boldsymbol\xi) 
\frac{\partial\Delta u_{i}}{\partial\xi_{\tau}} 
(\boldsymbol\xi,s) 
\nonumber\\&&\times
 (\nu_j\tau_{m} -\nu_m\tau_{j})
c_{ij pq}\frac{\partial G_{np}}{\partial \xi_q} ({\bf x},t;\boldsymbol\xi,s),
\label{eq:D4reduced2D}
\end{eqnarray}
where $\tau=\tau_1$ and $\eta=\eta_1$ are implied as in ordinary studies of two-dimensional problems.
%The third term of Eq.~(\ref{eq:D4reduced}) exactly vanishes in the two-dimensional problems without any additional assumptions.

%Although previous studies in three-dimensional problems have used the equation of motion for transverse waves (Fukuyama and Madariaga, 1998, Aochi and Fukuyama, 2001), it is reported that the resulting regularized expression consistently reproduces the two-dimensional results (Tada and Madariaga, 2001) derived from the equation of motion of the Green's function (\ref{eq:EOMGofff}) shown in Tada and Yamashita (1997). 

\subsubsection{Non-Hypersingular Stress Green's Functions}
%以上のように、正則化された積分方程式が得られた。
%..に対するカーネル、...に対するカーネルとみなすと、これらは積分の結果収束する関数になっている。

%愚直にを実行するとdozens termsが生じることが知られている()が、解析での困難な部分は大部分がテンソルに起因し、正則化自体はグリーン関数や弾性係数の一般的な性質だけから実行できるような比較的単純な構造を持っている。式第一項は..に対する応答を表す。第二項は、速度変化に対する慣性的な応答を表す。で、termsも名付ける.単純な意味でのdislocation にはなっていないとも明記.(面に沿ったスリップの勾配が生み出すのではないということが本質.というパンチライン.analytical rederivationのところでいう.)
%テンソル抜き去って解釈できない?

%また、これに..を適用することで、応力の表現として
%...
%を得る。
We have obtained the non-hypersingular integral equation, Eq.~(\ref{eq:D4reducedA}), for the displacement-gradient field. 
This also provides the non-hypersingular stress Green's function 
through Eq.~(\ref{eq:reexofstress}):
\begin{eqnarray}
&&\sigma_{kl}({\bf x},t) 
\nonumber\\&=&
c_{klmn}
\int^\infty_{-\infty} ds\int_{\Gamma(s)} d\Sigma(\boldsymbol\xi) \Delta \dot u_{i}(\boldsymbol\xi,s) \nu_m\rho \partial_s G_{ni}({\bf x},t;\boldsymbol\xi,s)
\nonumber\\&+&
\sum_\phi 
c_{klmn}c_{ij pq}
\int^\infty_{-\infty} ds\int_{\Gamma(s)} d\Sigma(\boldsymbol\xi) 
\frac{\partial \Delta u_{i}}
{\partial \xi_{\tau_\phi}}
\nonumber\\&\times&
(\nu_j\tau_{\phi m} -\nu_m\tau_{\phi j})
\frac{\partial G_{np}}{\partial \xi_q} ({\bf x},t;\boldsymbol\xi,s).
\label{eq:stressGreen}
\end{eqnarray}
%Eq.~(\ref{eq:derivoftorquishtensor}) can be used to evaluate the derivative of $(\nu_m\tau_{\phi j} -\nu_j\tau_{\phi m})$ as in Eq.~(\ref{eq:D4reduced}).

The corresponding two-dimensional expressions are obtained from the three-dimensional one given in Eq.~(\ref{eq:stressGreen}) both for the in-plane (modes I or II) and anti-plane (mode III) problems, 
just as the two-dimensional displacement-gradient Green's function, Eq.~(\ref{eq:D4reduced2D}), obtained from the three-dimensional one, Eq.~(\ref{eq:D4reducedA}). 

%
%なお、動的な場合には式..での項をいずれも同一の変数に対するカーネルとして、特に局所座標系が時間不変である場合に、
%\begin{eqnarray}
%..
%\end{eqnarray}
%という形に書き換えることができる。いずれもすべり速度に対する変数になっている。積分方程式の解析系の導出ではこちらが容易かもしれない。
%

We also obtain a non-hypersingular integral equation from Eq.~(\ref{eq:D4reducedA}) for the symmetric strain tensor $(\partial_m u_n+\partial_n u_m)/2$. 
%なお、式は書き下さないが、..+..およびからひずみテンソルは計算される。

\subsubsection{Non-Hypersingular Green's Functions in Static Problems}
Significant specializations of Eqs.~(\ref{eq:D4reducedA}) and (\ref{eq:stressGreen}) are 
the regularized displacement-gradient and stress Green's functions for the static problems. They are obtained in the quasi-static limit where the slip $\Delta u$  
is treated as time invariant: 
\begin{eqnarray}
&&\partial_{m} u_n({\bf x}) 
=
\sum_\phi 
c_{ij pq}
\int_{\Gamma} d\Sigma(\boldsymbol\xi) 
\frac{\partial \Delta u_{i}}
{\partial \xi_{\tau_\phi}}
(\boldsymbol\xi)  
\nonumber\\&&
\times
(\nu_j\tau_{\phi m} -\nu_m\tau_{\phi j})
\frac{\partial G_{st,np}}{\partial \xi_q} ({\bf x};\boldsymbol\xi)
\label{eq:displacementgradientGreenstatic}
\\
&&
\sigma_{kl}({\bf x}) =
\sum_\phi 
c_{klmn}c_{ij pq}
\int_{\Gamma} d\Sigma(\boldsymbol\xi) 
\frac{\partial \Delta u_{i}}
{\partial \xi_{\tau_\phi}}
(\boldsymbol\xi)  
\nonumber\\&&
\times
(\nu_j\tau_{\phi m} -\nu_m\tau_{\phi j})
\frac{\partial G_{st,np}}{\partial \xi_q} ({\bf x};\boldsymbol\xi),
\label{eq:stressGreenstatic}
\end{eqnarray}
where ${\bf G}_{st}=\int ds {\bf G}(=\int dt {\bf G})$ is the static Green's function, independent of the time $t$, and we have eliminated the time dependences of the displacement gradient and stress from the left-hand sides.
The inertial contribution, represented by the first terms in Eqs.~(\ref{eq:D4reducedA}) and (\ref{eq:stressGreen}), vanishes in the quasi-static limit. 

Two-dimensional expressions are obtained from Eqs.~(\ref{eq:displacementgradientGreenstatic}) and (\ref{eq:stressGreenstatic}) in the same way as in the dynamic cases. 

The derivatives of the Green's function can be found in \citet{okada1992internal,tada2000non} for three-dimensional isotropic media, and they are listed in Appendix \ref{sec:A} for the two-dimensional ones. 
%静問題は、$\Delta u$を時間不変とする極限で得られる。時間微分をの極限として、...を得る。

\section{
The Paradox of Smooth and Abrupt Bends Revisited
}\label{sec:4}
We have investigated carefully the development of the non-hypersingular stress Green's function, Eq.~(\ref{eq:stressGreen}), from the general homogeneous Green's function. 
As an application, we reconsider below the paradox of smooth and abrupt bends. 
In \S\ref{sec:41} with using the quasi-static limit, Eq.~(\ref{eq:stressGreenstatic}), of Eq.~(\ref{eq:stressGreen}) that we derived, 
we treat the problem of uniform shear for which \citet{tada1996paradox} showed that the stress analytically must be zero in the previous formulations. 
In \S\ref{sec:42}, we show that the differential geometry on the curved boundary sheds light on the root of the paradox.

\subsection{Numerical Test of Dislocation Problems on Curved Fault Geometry}\label{sec:41}
The paradox of smooth and abrupt bends was recognized in an investigation of the stress fields caused by the slip on a curved fault geometry~\citep{tada1996paradox}. The geometry studied was constructed by an arc connecting two infinitely long straight half-lines in two-dimensional space. Fig.~\ref{fig:1} shows this geometry, where the arc has the curvature radius $R$. We here fix the arc length to $\pi R/4$. By solving the elastic problem for such a curved fault by using the stress Green's function of \citet{tada1996boundary,tada1997non} for an isotropic, homogeneous, elastic medium, \citet{tada1996paradox} reported that the stress fields in the discretized cases are different from those in the un-discretized case, even in the limit ($\Delta \xi/R\to 0$) of an infinitesimal discretization length $\Delta \xi$.

The clearest explanation of the paradox is provided in \citet{tada1996paradox} by using a case of constant shear slip $\Delta {\bf u}=\Delta u_0{\bf t}(\boldsymbol\xi)$ along a smoothly curved fault (a smooth curve) for an arbitrary non-zero constant $\Delta u_0$. 
They solved this case analytically based on \citet{tada1996boundary} or equivalently \citet{tada1997non}. The solution predicts zero stress over the entire medium. Despite this analytical prediction, their result for a discretized fault (a line chain) exhibited the finite stress. 

Here we revisit the paradox of smooth and abrupt bends with this constant shear-slip problem in the curved geometry.
To confirm the validity of the obtained result, we first study the stress field off the fault (the off-fault stress) using Eq.~(\ref{eq:stressGreenstatic}). In this setting, we can use the hypersingular expression for the stress field calculated from Eq.~(\ref{eq:displacementgradienthom}) (in the quasi-static limit) using Eq.~(\ref{eq:reexofstress}). We later test the stress on the fault (the on-fault stress) for a variable slip distribution.

In the following numerical test, we compare the non-hypersingular ones we obtained [Eq.~(\ref{eq:stressGreenstatic})] and those previously obtained~\citep{tada1996boundary,tada1997non}. 
We also compute 
i) the line chain, which is a fundamental tool in a large part of numerical analysis, and ii) the hypersingular stress Green's function obtained from Eq.~(\ref{eq:displacementgradienthom}) with Eq.~(\ref{eq:reexofstress}).
The hypersingular expression gives the correct answer, as long as it is applied to evaluate the off-fault stress, and 
because of its simple derivation [it is just a derivative of the representation theorem, Eq.~(\ref{eq:Bettihom})], it can serve as a reference for the correct value of the off-fault stress. 
That is, the correct non-hypersingular expressions need to give the same off-fault stress as the hypersingular one. 

We compute or calculate these values in the following manner.
The derivative of the tangential vector of the slip in Eq.~(\ref{eq:stressGreenstatic}) is computed with using
\begin{eqnarray}
\partial_{\tau} \boldsymbol\tau
&=& 
\frac {\partial\boldsymbol\tau}{\partial\eta}
=
\kappa_\phi\boldsymbol \nu 
\label{eq:derivtau}
\\
\partial_{\tau}\boldsymbol \nu &=&\frac {\partial\boldsymbol \nu}{\partial\eta}=-\kappa_\phi\boldsymbol \tau_\phi
\label{eq:derivnu}
\end{eqnarray}
where the coefficient $\kappa$ is called the curvature (the inverse of which is called the curvature radius)~\citep{pressley2010elementary}; it is given by %denoted by $\kappa_\phi (\boldsymbol\xi,s)$, is given in the form
\begin{eqnarray}
\kappa:=\frac{\partial\boldsymbol\tau}{\partial\eta}\boldsymbol\nu^{-1}.
\end{eqnarray}
Eqs.~(\ref{eq:derivtau}) and (\ref{eq:derivnu}) are obtained 
by approximating the geometry in an infinitesimally small space around each location by a circular arc 
through 
%After projected onto a plane spanned by $\boldsymbol\tau_\phi$ and $\boldsymbol\nu$,
%the inverse of the local radius (the curvature radius) called
Eq.~(\ref{eq:curvatureproptau}) and Eq.~(\ref{eq:curvaturepropnu}) and replacing $R^{-1}$ by $\kappa$. 
In Fig.~\ref{fig:1}, $\kappa=1/R$ on the arc and $\kappa=0$ on the half-lines.
%hold in general smoothly curved faults for given $\tau=\tau_\phi$ and $\eta=\eta_\phi$. 
%That is, the infinitesimal spatial changes of the local coordinate axes is given in the differential forms, 
%局所座標を貼るベクトルの局所的な移り変わりは、$\kappa_\phi (\xi)$は$\tau_\phi , \nu$で貼られる面に射影された$\xi$近傍の断層形状での曲率によって表される。
%The left tangent axis does not change in the two-dimensional geometry,
%\begin{eqnarray}
%\partial_{\tau_\phi} \boldsymbol\tau_{\bar \phi}=0, 
%\end{eqnarray}
%where $\bar \phi$ denotes the one of 1,2 being not $\phi$ ($\phi=1,\bar\phi=2$ or $\phi=2,\bar \phi=1$).
We calculate the derivatives of the Green's function ${\bf G}$ in Eqs.~(\ref{eq:stressGreenstatic}) and (\ref{eq:displacementgradienthom}) analytically, as shown in Appendix \ref{sec:A}. 
We used Simpson's rule or the double-exponential scheme for the numerical integrations. The integrated values of the off-fault stress are independent of the numerical integration schemes to within the numerical precision. 
The on-fault stress is integrated in the Cauchy sense with the double-exponential scheme~\citep{mori2001double} of \citet{ooura1999robust}.
The line-chain solution can be found in \citet{ando2007efficient} as the quasi-static limit of the result in \citet{tada2001dynamic}, which is a discretization of the previous non-hypersingular expression of \citet{tada1997non}.
The line chain constitutes the polygonal lines inscribed within the original curve. 
Note that the value of the previously obtained stress Green's function is exactly zero~\citep{tada1996paradox}. 

\begin{figure}%[tbp]
  \includegraphics[width=87mm]{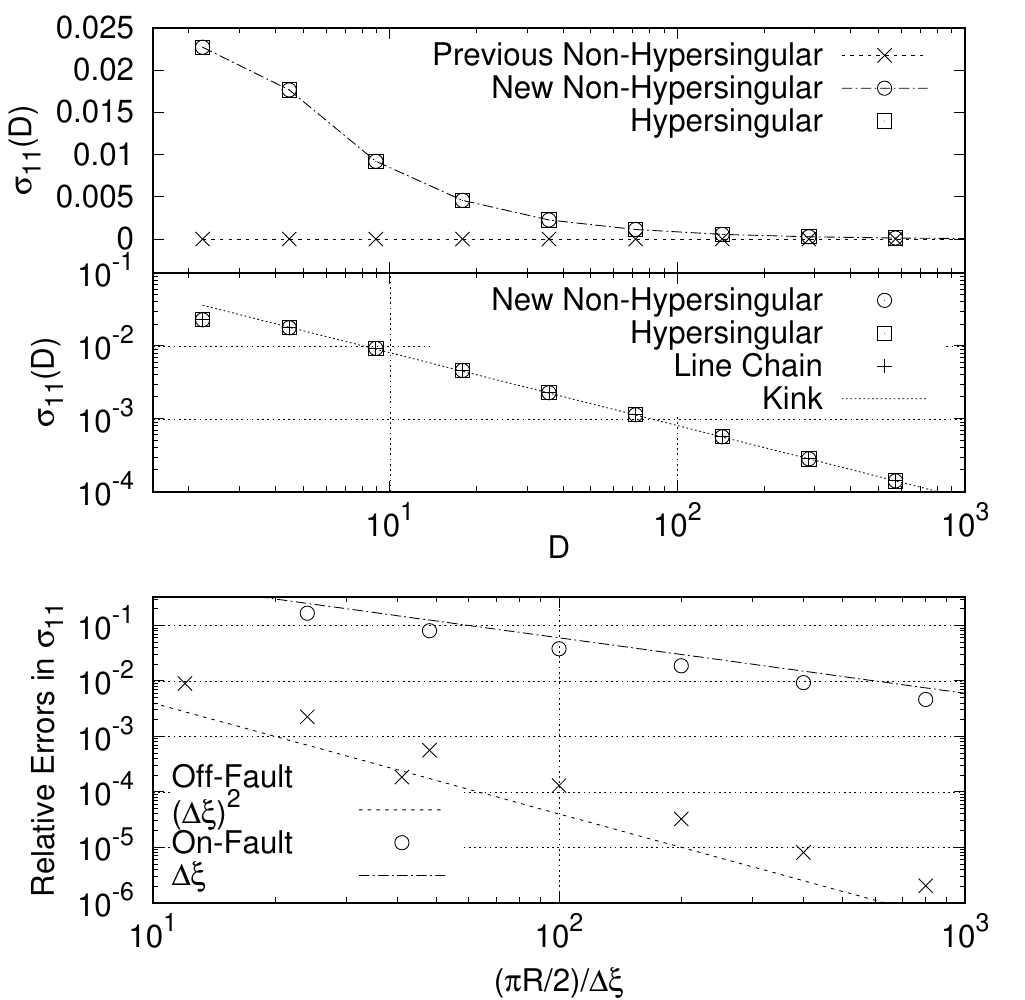}
 \caption{
Numerical comparisons of the stress Green's functions, detailed in the text of \S\ref{sec:41}.
(Top) Predicted values of $\sigma_{11}$ along the stress evaluation line in Fig.~\ref{fig:1}. The bottom part of the panel shows the results in the log scale for further investigation of the results shown in the top part. 
(Bottom) Element-size $\Delta \xi$ dependence of the relative errors in $\sigma_{11}$ for the line-chain solution, which converges to the un-discretized one. %The number of elements is inversely proportional to the mesh-size $\Delta \xi$, proportional to the number of elements on the arc. 
%The off-fault case is evaluated at $D=2.236$ on the stress evaluation line with constant shear slip. 
% by a case corresponding to the top bottom panel ($\sigma_{11}(D=2.236)$). 
%On-fault one is at the midpoint of the element covering the right edge of the arc with $\Delta u_\tau=\sqrt{(\pi R/4)^2-\eta^2}H((\pi R/4)^2-\eta^2)$. 
The asymptotes $\mathcal O((\Delta \xi)^2)$ and ($\mathcal O(\Delta \xi)$) 
are indicated by dotted lines in the panel, respectively, for the off-fault stress caused by a constant slip and for the on-fault due to a variable slip. 
%$??$The following (the supplemental information for the off-fault case) will be checked and modified...with varying slip; the slip is assigned only in the arc and is proportional to the square of the distance from a closer arc edge. 
%(Right) summary of the left top panel.
 }
\label{fig:2}
\end{figure}

Fig.~\ref{fig:2} (top) shows the result for the 1,1-component $\sigma_{11}$ of the stress tensor for each distance $D$ along the specific line shown in Fig.~\ref{fig:1}. We obtained the plotted numerical values for the hypersingular/non-hypersingular expressions with sufficient accuracy. The angle parameter $a$ in Fig.~\ref{fig:1} is set at $a=\mbox{atan}(2)-\pi/4$ in the simulation. The result is normalized by taking the unit $R\to1$, and the rigidity is 1. The Poisson ratio is set at 1/4. 

It is quite remarkable that a non-zero value is predicted by the non-hypersingular stress Green's function we obtained (labeled ``New Non-Hypersingular'' in Fig.~\ref{fig:2}). This result is clearly different from the prediction of the previously obtained hypersingular stress Green's function (labeled ``Previous Non-Hypersingular'' in Fig.~\ref{fig:2}). 
The result of the hypersingular one (labeled ``Hypersingular'' in Fig.~\ref{fig:2}) also coincides admirably with our non-hypersingular one, and supports our non-hypersingular expression. 

Furthermore, perhaps surprisingly, the result for the line chain also coincides with the result of the hypersingular stress Green's function and our non-hypersingular one. This suggests that the correct answer can be obtained from a line chain that discretizes a smooth fault into multiple short lines. Given the zero value of the previous non-hypersingular stress Green's function on the smooth curve, this implication can be paraphrased as indicating that the previously obtained non-hypersingular stress Green's functions are correct only on a discretized flat boundary and that they are not suitable for treating a smooth curve directly. 
%Only the result interpolated on the line chain (on the flat boundary) seems to have been correct in the realm of the previously obtained stress Green's function. 

In addition, the three computational results--except for the previously obtained stress Green's function (on a smooth curve)--converged to the result of a kinked fault (corresponding to the limit $R\to0$, substantially $R/D\to 0$) as the distance $D$ from the kink becomes larger. These non-zero values are hence consistent with the property of elastic equations~\citep{aki2002quantitative} that, as a receiver becomes distanced from a source, the stress caused by a source approaches to that caused by a point source which gives the same total amount of the dislocation.

Let us clarify the relation between the non-hypersingular stress Green's function we obtained and the prediction of the line chain. As detailed later, our non-hypersingular expression coincides with the previous non-hypersingular one for a flat boundary. Hence the line-chain result, obtained from the previous non-hypersingular expression of using the discretized flat boundary elements, is also the discretization of our non-hypersingular one; this is also surprising because our and previous results are clearly different when we consider the original result without any discretization of the boundary.  

Fig.~\ref{fig:2} (bottom) shows the differences in the line-chain result and our non-hypersingular expression for the off-fault stress and on-fault stress. As the discretization length $\Delta \xi$ of the boundary element becomes shorter, the line-chain result converges to our non-hypersingular result within the error of $\mathcal O((\Delta \xi)^2)$; the plotted result is obtained for $D=2.236$. On the other hand, one of the most demanding cases--the on-fault stress with a variable slip distribution--converged to our non-hypersingular result to within the error of $\mathcal O(\Delta \xi)$; we evaluated the on-fault stress at $\eta=8/\pi$ with an example $\Delta u_\tau=\cos^2(2\eta/R)H(\pi R/4-|\eta|)$ of slip distributions that produce H\"{o}lder-continuous dislocations $\partial_\tau\Delta u$ over the fault.
These results are consistent with the analytical error estimates for the discretized solution of our non-hypersingular stress Green's function (Appendix \ref{sec:B}). 

As above, as long as one uses our non-hypersingular stress Green's function, the plausibility of which is supported by the hypersingular one, 
modeling with a discretized boundary can reproduce the result expected from the original smooth curve. This means that the paradox of smooth and abrupt bends does not exist for the correctly modified, non-hypersingular stress Green's function. 
In addition, as the previous expressions are correct only for the flat cases, 
the root of the paradox--which was missed in several previous studies--must be some geometrical factors that only appear on the curved faults. 
%
%
%Before considering the reason of the inconsistency between the representation theorem with the previous non-hypersingular expression on the smootly curved fault, as in the off-fault case, we test the convergence of on-fault stress computed by the line-chain result to the non-hypersingular stress Green's function (Eq.~(\ref{eq:stressGreenstatic})) we obtained. 
%The result is shown in Fig.~\ref{fig:2} plotting the dependence of the solution error on the mesh-size  $\Delta \xi$ of the element both in the off-fault and off-fault cases; $D=2.34$ is chosen in the result and the on-fault stress is evaluated at the midpoint of the element covering the right edge of the arc.
%In both cases, as $\Delta \xi$ becomes small, the solutions of the line chain converge to the non-hypersingular stress Green's function (Eq.~(\ref{eq:stressGreenstatic})) of us. 
%The errrors are of $\mathcal O(\Delta \xi)$ in the on-fault case, while $\mathcal O((\Delta \xi)^2)$ in the off-fault case. As mentioned in the appendix A, these error orders holds in wide classes of the solutions. 
%Given the results of Fig.~\ref{fig:2}, the line chain result is considered to be plausible.  
%
%
%The above results are summarized as follows. The hypersingular stress Green's function, directly obtained from the representation theorem, supported our non-hypersingular one and rejected the previous non-hypersingular one on the curve boundary. 
%Nevertheless, the line chain result, obtained from the previous non-hypersingular expression on a planar boundary, is supported from our non-hypersingular one. 

\subsection{
Differential Geometry on Curved Faults Missed in Previous Non-Hypersingular Expressions
}\label{sec:42}
Numerical experiments suggest that the paradox is caused by the previously obtained non-hypersingular stress Green's function that erroneously works on the smoothly curved faults.  
%Mode II Dislocation とか Mode III Dislocationとして解釈することはできないというのが本質。この点は、Pierreらの論文で扱われると。他方、line chains上のmode II, mode III dislocationsとしてnumericalに扱うことは可能.Mode III slipやMode II slipとして扱うことは可能.
%Surprisingly, they showed the same values.
%\footnote{
%This had been done to persuade Pierre to admit the wrongness of Tada and Yamashita, but I also had made a mistake when I had executed this computation; I had supposed the piecewise-constnat solution is wrong, since I made a mistake when evaluate the contriubtion from the junction (shown in Fig.1, right panel). Now I recognize the correctness of the piecewise-constant solution.}
In retrospect, \citet{tada1996paradox} showed that the paradox is essentially caused by
the non-uniqueness of the zero-stress field;
by using the stress Green's function of \citet{tada1996boundary},
they showed that 
the stress-field becomes zero over the entire medium 
 when either of the following two conditions is satisfied:
\begin{eqnarray}
\forall (a,b),\,
\partial_a ^{(\xi)} \Delta u_b=0 ,
\label{eq:zerostress}
\end{eqnarray}
%except accidental stress vanishment because of some symmetry. 
or 
\begin{eqnarray}
\partial^{(\xi)} \Delta u_\tau=\partial^{(\xi)} \Delta u_\nu= 0.
\label{eq:zerostressnot}
\end{eqnarray}
However, this conclusion misses the change in the local coordinate axes, Eqs.~(\ref{eq:derivtau}) and (\ref{eq:derivnu}). 
We already know that the two conditions above are not equivalent, given Eqs.~(\ref{eq:derivtau}) and (\ref{eq:derivnu}), and their differences can be written in the following forms: 
\begin{eqnarray}
\partial^{(\xi)} (\Delta u_\tau \tau_b )
&=&
\tau_b \partial^{(\xi)} \Delta u_\tau 
+
 \nu_b\Delta u_\tau \kappa
\label{eq:entrancetoPSA1}
\\
\partial^{(\xi)} (\Delta u_\nu \nu_b )
&=&
\nu_b \partial^{(\xi)} \Delta u_\nu
-
 \tau_b\Delta u_\nu \kappa
\label{eq:entrancetoPSA2}
\end{eqnarray}
%where $\kappa$ is the curvature (the inverse of the curvature radius).
The second terms in these equations, which are proportional to the curvature, are what have been neglected in claiming the abovementioned equivalence.
%, called the curvature term by Pierre.
Indeed, the terms proportional to $\kappa$ are not included in the starting-point equation of \citet{tada1996paradox} [Eq. (1) in their paper] as in other previous studies.
As noted from Eq.~(\ref{eq:D4reducedA})--or equivalently from Eq.~(\ref{eq:D4reduced2D}) for two-dimensional cases--zero stress over the entire medium occurs only when Eq.~(\ref{eq:zerostress}) is satisfied. Eq.~(\ref{eq:zerostressnot}) does not necessarily mean that the stress field is zero throughout the entire medium. 
%Furthermore, as noticed from Eq.~(\ref{eq:D4reducedA}), the general condition for the zero stress over the entire medium seems even not to be Eq.~(\ref{eq:zerostress}) and is expressed as
%\begin{eqnarray}
%\forall (\phi,i,j,m),
%\,
%\partial^{(\xi)}_\phi[\Delta u_i(\nu_{j}\tau_{\phi m}-\nu_{m}\tau_{\phi j})]=0.
%\label{eq:zerostressgen}
%\end{eqnarray}
%The difference between Eqs.~(\ref{eq:zerostressgen}) and (\ref{eq:zerostressgen}) is needed to be cared only in the three-dimensional problems, as mentioned when Eq.~(\ref{eq:D4reduced}) is derived.

The root of the paradox is as above caused by
the differentials of the local coordinate axes, which were missed in previous studies 
and which are not contained in our stress Green's function, as shown by the numerical results.
This difference of the new and previous non-hypersingular expressions indeed leads to the new finding detailed in our companion paper (Romanet, Sato, and Ando), which will give a physical implication concerning the curvature.  

The condition $\partial^{(\xi)}_a \Delta u_b=0$ is equivalent to $\partial^{(\xi)}_a \Delta u_{\tau_\phi}=0$ only for the anti-plane problems. It is also consistent with the report by \citet{tada1996paradox} that the paradox vanishes in the anti-plane problems.

In addition, the second terms in Eqs.~(\ref{eq:entrancetoPSA1}) and (\ref{eq:entrancetoPSA2}) vanish for the line chain, and then our and previous expressions of the non-hypersingular stress Green's function are equivalent for flat boundary sources. 
Moreover, in our non-hypersingular expression, 
the interpolated slip, with the piecewise-constant function in the global coordinate system, gives exactly the same stress field as that of the piecewise-constant slip on the line chain inscribed within the original curve, without requiring any discretization of the fault geometry (mentioned in Appendix \ref{sec:B}).
These results actually provide a logical reason why the line-chain solution obtained from the previously derived non-hypersingular Green's function converges to our non-hypersingular solution in the numerical experiment shown earlier. 
This equivalence of the solutions also holds for three-dimensional problems of smooth curve and subdivided flat patches. 

%The terms neglected by Tada and Yamashita's analysis become clear when we take a projection of dislocation onto the fault in eq. (\ref{eq:gennonhypersingulareq}) 
%\begin{eqnarray}
%\partial_\phi^{(\xi)} (\Delta u_b )
%&=&
%\partial_\phi^{(\xi)} 
%(\tau_{\phi b}\Delta u_{\tau_\phi}
%+\nu_b\Delta u_\nu
%+\tau_{\bar \phi b}\Delta u_{\tau_{\bar \phi}}
%)
%\\
%&=&
%\tau_{\phi b}
%\partial_\phi^{(\xi)} 
%\Delta u_{\tau_\phi}
%+\nu_b
%\partial_\phi^{(\xi)} 
%\Delta u_\nu
%+\tau_{\bar \phi b}
%\partial_\phi^{(\xi)} 
%\Delta u_{\tau_{\bar \phi}}
%\nonumber
%\\&&
%+\kappa_{n\phi}[
%\nu_b
%\Delta u_{\tau_\phi}
%-
%\tau_{\phi b}
%\Delta u_\nu
%]
%+\kappa_{g\phi}[
%\tau_{\bar \phi b}
%\Delta u_{\phi}
%-
%\tau_{\phi b}
%\Delta u_{\bar\phi}
%]
%+\kappa_{r\phi}[
%\nu_b
%\Delta u_{\tau_{\bar \phi}}
%-
%\tau_{\bar \phi b}
%\Delta u_\nu
%],
%\end{eqnarray}
%where $\bar\phi(=1,2)$ is the other number from $\phi$, 
%and
%$\kappa_{n\phi},\kappa_{g\phi},\kappa_{r\phi}$
%are the normal curvature, geodesic curvature, and relative torsion, respetively, for the $\phi$-th shear direction.

\section{Discussion}
Our analytical and numerical investigations of the stress Green's function have resolved a previously identified problem (the paradox of smooth and abrupt bends) concerning the convergence of discretized solutions to the true solution. We have shown that a solution for discrete segmented geometry converges to the corrected un-discretized solution in the limit of sufficiently fine discrete elements. In that sense, the smooth and discretized/segmented faults are equivalent, and therefore can be used without special discrimination. 
This may be the answer to the above-mentioned ambiguity concerning the adequacy of smooth and discretized faults in modeling real faults~\citep{duan2005multicycle,kase2006spontaneous}. 
Previous studies using discretized faults (e.g., Aochi \& Fukuyama 2002) are verified in that way. 
Given that boundary geometries are discretized with segments (subdivided flat patches) in most of numerical studies~\citep[e.g.,][]{aochi2002three} even for smooth boundaries and even when using finite-element/difference methods~\citep[e.g.,][]{oglesby20031999}, our result is finally to show the consistency between the numerical modeling and analytical studies of smooth (undiscretized) faults. 
It is paired with another implication that fine segments can be modeled by smooth curves in the description of macroscopic stress field as has been done in the tectonic modeling.

We eventually found that the cause of the apparent paradox is due to the spatial changes in the axes of the local coordinates, which had been missed in the previous studies that considered non-planar geometries~\citep[e.g.,][]{jeyakumaran1994curved,tada1996boundary,aochi2000spontaneous,tada2006stress}. 
Our finding is a negative resolution of the paradox, in the sense that previous analytical studies missed the differentiation of the local coordinate axes and hence obtained inconsequent results for the given problems.
This theoretical oversight is mathematically simple and is corrected 
%by Eqs.~(\ref{eq:derivtau3D}) and (\ref{eq:derivnu3D}) (
from the differential geometry of the local coordinate system, as in Eqs.~(\ref{eq:derivtau}) and (\ref{eq:derivnu}). Nevertheless, this will be a missing link that leads to the nontrivial physical suggestion that the stress field induced by a quasi-statically imposed slip of respective modes ($\Delta u_\nu$, $\Delta u_{\tau_1}$, $\Delta u_{\tau_2}$) cannot be described only with the differentials of the slip along the fault 
($\partial_{\tau_{\phi}}\Delta u_\nu$, $\partial_{\tau_{\phi}}\Delta u_{\tau_1}$, $\partial_{\tau_{\phi}}\Delta u_{\tau_2}$), i.e., the dislocation.
This point seems not to have been found from the discretized numerical analyses, and as clarified in our companion paper (Romanet, Sato, and Ando), this stress gap is compensated by a hidden mechanics to relate the stress with the nonplanarity of the geometry.

Our derivation relies on the spatiotemporal translational symmetry of the Green's function [Eqs.~(\ref{eq:spatialtrans}) and (\ref{eq:temporaltrans}), used to obtain Eqs. ~(\ref{eq:D4hom}) and (\ref{eq:projEOMGofff}), respectively], where the location and time dependence of Green's function is fully described by the relative location and time (${\bf x}-\boldsymbol \xi$, $t-\tau$) between a receiver and source. The derivation of our non-hypersingular stress Green's functions are then limited by the applicability of Eqs.~(\ref{eq:spatialtrans}) and (\ref{eq:temporaltrans}), which hold only in the time-invariant homogeneous medium of full space. Other relations used in the derivation such as the equation of motion, reciprocities, representation theorem, and Stokes' theorem hold in arbitrary elastic media, and thus the technical difficulty of our derivation in inhomogeneous media is purely due to the requirements of spatiotemporal translational symmetries. 
For the same reason, an analytic derivation (Appendix \ref{sec:B}) is limited to the homogeneous case to show the convergence of the discretized solution to an original solution.
Nonetheless, we can treat the inhomogeneous medium with the Green's function of homogeneous media, by partitioning a whole medium into a sufficiently fine subset of effectively homogeneous media (multiregion approaches)~\citep[p301]{bonnet1999boundary}. This indirect extension of our results via multiregion approaches assures the generality of the equivalence between the smooth and infinitesimally discretized faults shown in this study, and the physical insight of our companion paper in the inhomogeneous media. The implication of our results is as above free from such a technical issue of our derivation of the non-hypersingular stress Green's functions. 

\begin{figure}%[tbp]
  \includegraphics[width=100mm]{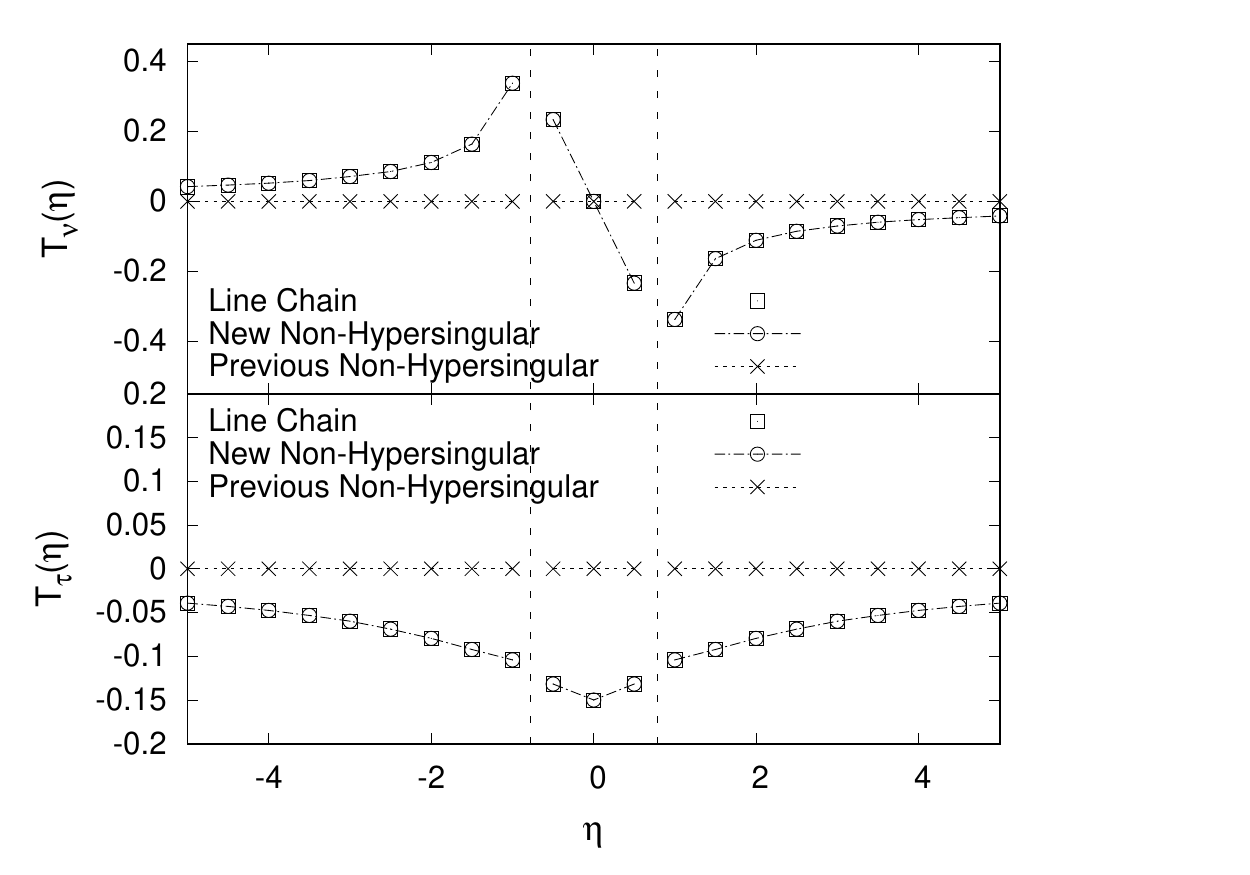}
 \caption{
Traction fields caused by a constant shear slip. 
The computational details are given in the discussion section. 
The vertical dotted lines indicate the locations at which the circle connects to the straight-line segments.
 }
\label{fig:4}
\end{figure}

As for the stress field, the (on-fault) traction field of a line chain is consistent with our non-hypersingular expression (Fig.~\ref{fig:4}); the slip in this figure is set to be the same constant shear slip as in the off-fault stress case considered in the numerical experiments. Although \citet{tada1996paradox} showed that the smooth curve and line chain give normal tractions of different signs in a problem with a stress boundary condition (Fig.~2 of their paper), this discrepancy is apparent and faultily arises because of their smooth curve result missing the term proportional to the curvature in Eq.~(\ref{eq:entrancetoPSA1}). Note that the normal traction in the smooth-curve result of \citet{tada1996paradox} is spatially variable due to the stress boundary condition they imposed. Indeed, the normal traction in the smooth-curve result becomes zero when \citet{tada1996paradox} consider the boundary condition of the constant shear/normal slip, as they themselves pointed out (the text accompanying Fig.~3 in their paper). We treat the crack problem further in the companion paper.

%In three-dimensional problems, we may need to further care the derivative $\partial_{\tau_\phi} (\nu_j\tau_{\phi m} -\nu_m\tau_{\phi j})$ as shown in Eq.~(\ref{eq:D4reduced}). 
Our non-hypersingular Green's functions are actually equivalent to those obtained by \citet{bonnet1999boundary}, where his Eq.~(4.24) (p75) is for elastostatics and Eq.~(7.61) (p176) for elastodynamics. 
For example, 
our Eq.~(\ref{eq:stressGreen}) can be expressed as 
\begin{eqnarray}
&&\sigma_{ij}({\bf x},t)
\nonumber\\
&&= 
-\int_{-\infty}^\infty ds \int_\Gamma d\Sigma(\boldsymbol\xi) 
\rho c_{ijkl}G_{ka}({\bf x},t;\boldsymbol\xi,s)
\partial_s\Delta \dot u_a (\boldsymbol\xi,s)
\nu_l
\nonumber\\&&+
\int_{-\infty}^\infty ds \int_\Gamma d\Sigma(\boldsymbol\xi) c_{ijkl}\Sigma^k_{ab}({\bf x},t;\boldsymbol\xi,s)
D_{lb}\Delta u_a(\boldsymbol\xi,s)
\end{eqnarray}
in the nomenclature of \citet{bonnet1999boundary}, with $D_{lb}:=-\sum_\phi(\nu_l\tau_{\phi b}-\nu_b\tau_{\phi l})\partial_{\tau_\phi}$ and $\Sigma^k_{ab}({\bf x},t;\boldsymbol\xi,s)
:=c_{abcd}\partial_c G_{dk}({\bf x},t;\boldsymbol\xi,s)$. Note that the reciprocal Green's function $G_{ab}(\boldsymbol\xi,s;{\bf x},t)$ is used with single-force/traction contributions for finite-spaced media in the original expression of \citet{bonnet1999boundary} as well as with the outward normal vector. 
The above expressions of \citet{bonnet1999boundary} will be shown explicitly to be identical with our Eq.~(\ref{eq:stressGreen}) in the companion paper, by following the derivation of \citet{bonnet1999boundary} being different from conventional local-coordinate techniques we employed.
Our finding may thus be just to distinguish the expression given by \citet{bonnet1999boundary} from previously obtained alternatives that erroneously miss the derivatives of the local coordinate axes. 
Nevertheless, this difference among these previous expressions was previously unrecognized and at least our careful exploration of the correct non-hypersingular expressions has successfully revealed such differences and clarified which of the expressions is correct.  

As above, even for static problems, the stress source is not the extensively studied pure dislocation due to the differentials of the local coordinate axes. It may be a consolation for theorists that such a source can still be safely reduced to a dislocation through discretization using flat boundary elements, as shown by our line-chain results. 
Since the stress Green's function, Eq.~(\ref{eq:stressGreen}), is a sum of non-hypersingular integral equations convolving the slip rate or the slip gradient along the fault, it can be evaluated as a Cauchy integral as long as H\"{o}lder continuity is satisfied by the slip rate and slip gradient, in addition to the differentiability of the slip assumed in the derivation. 
As frequently done in fracture mechanics,  
Eq.~(\ref{eq:stressGreen})
is also integrable in the Cauchy sense, even for a piecewise-constant slip rate, so long as the receiver is forbidden to be located at the edge of the interpolation region, where the gradients of $\Delta u_i$ diverge.
In that case, the intractable non-hypersingularity of Eq.~(\ref{eq:displacementgradienthom}) is regularized as  radiation damping in the integral equation for the slip rate (the inertial term) in Eq.~(\ref{eq:D4reducedA}), as pointed out by~\citet{geubelle1995spectral,fukuyama1998rupture}. 
By the same logic, the quasi-static expression, Eq.~(\ref{eq:displacementgradientGreenstatic}), regularized for a H\"{o}lder-continuous dislocation, can be 
evaluated as a Cauchy integral even for piecewise-constant slip, except for the case in which the receiver is located at the edge of the interpolation region. %, where the extracted hypersingularity (the radiation damping proportional to the slip rate) is cancelled. It is not obvious whether non-isotropic cases keep such heuristic non-hypersingularity of Eq.~(\ref{eq:D4reduced}) and (\ref{eq:displacementgradientGreenstatic}) for the piecewise-constant slip rate. 

%Eq.~(\ref{eq:D4reduced}) can be discretized in a variational approaches as in~\citet{nedelec1982integral}, although we imposed the piecewise-constant interpolation, and the mid-point collocation for the on-fault stress evaluation. This will give an alternative variational way of treating elastic boundary problems, basically treated by Kelvin's fundamental solution~\citep{nishimura1989regularized,bonnet1995regularized}. 

\section{Conclusion}
We have examined a paradox reported by \citet{tada1996paradox} that a solution with the discretization for a  dislocation/crack problem does not converge to the original solution on a curved boundary even in the continuous limit. We first develop a compact alternative expression for the non-hypersingular stress Green's function. 
Second we test the various stress Green's functions numerically. 
The numerical experiment refutes the paradox and suggests that 
the analyses of \citet{tada1996boundary} and subsequent studies are inadequate for a smoothly curved boundary. The numerical experiment further clarifies that such inadequate previous formulation of a smooth curve converges to the corrected solution after discretization with flat boundary elements. Appended analysis explains this convergence as a geometry-independent property of the corrected non-hypersingular stress Green's function. 
By examining the spurious zero-stress field condition, which was pronounced as the root of the paradox by \citet{tada1996paradox}, 
we have found that previous studies missed the differentiation of the local coordinate axes.  
%the paradox of curved faults is theoretically and numerically rejected. 
These results consistently show the equivalence between the smooth and infinitesimally segmented fault geometries.
Their equivalence reconciles the analytical methods for smooth curved boundaries and numerical methods using discretized flat patches, and the distinctive mechanics of non-planar faults, hidden behind such equivalence, will be clarified in our companion paper (Romanet, Sato, and Ando).

%\newpage 

\begin{acknowledgments}
We thank Dr. Bonnet for valuable discussions.
We also thank Dr. Eric Dunham and anonymous reviewer who helped improve the manuscript.
This work was supported by JSPS KAKENHI 18KK0095 and 19K04031.
D.S. was supported by MEXT KAKENHI Grant Numbers JP26109007.
P.R. received support from KAKENHI 16H02219.
\end{acknowledgments}
\section*{AUTHOR CONTRIBUTION STATEMENT}
D.S. wrote the manuscript. D.S. initially found the regularization technics of the boundary-integral equation. P.R. found the effect of curvature and the implications in 2D. Both D.S. and P.R. partici- pated in combining the two works and discussed the results. R.A. initiated the project and found the mistake in the work of \citet{tada1997non}. Finally, all the three authors have read and ap- proved the present manuscript.
\bibliographystyle{gji}
%\bibliography{reference_SPA.bib}

\appendix

\section{Derivatives of Green's Functions}
\label{sec:A}
In the isotropic cases $c_{ijkl}=\lambda\delta_{ij}\delta_{kl}+\mu(\delta_{ik}\delta_{jl}+\delta_{il}\delta_{jk})$,
two-dimensional expressions of the static Green's function ${\bf G}_{st}$ are for example found in 
\citet{maruyama1966two,tada1997non} in the forms,
\begin{eqnarray}
G_{st,33}({\bf x};\boldsymbol\xi)&=&\frac 1 {2\pi\mu}(-\log r)
\\
G_{st,ij}({\bf x};\boldsymbol\xi)&=&\frac 1 {4\pi\mu}
[\gamma_i\gamma_j(1-p^2)-\delta_{ij}(1+p^2)\log r
]
\end{eqnarray}
for $i,j=1,2$
with $r:=|{\bf x}-\boldsymbol\xi|$, $\gamma_i=(x_i-\xi_i)/r$ and $p=\sqrt{\mu/(\lambda+2\mu)}$, where $\lambda$ and $\mu$ are respectively Lame's first and second parameters.
By using $\partial_i^{(\xi)}(x_j-\xi_j)=-\delta_{ij}$ and $\partial_i^{(\xi)}\gamma_j=-(\delta_{ij}-\gamma_i\gamma_j)/r$,
we get the derivatives of ${\bf G}_{st}$ in two-dimensional problems as 
\begin{eqnarray}
\frac{\partial G_{st,33}}
{\partial \xi_i}
({\bf x};\boldsymbol\xi)&=&\frac 1 {2\pi\mu}\frac{\gamma_i} r
\\
\frac{\partial G_{st,ij}}
{\partial \xi_k}
({\bf x};\boldsymbol\xi)&=&-\frac 1 {4\pi\mu}
\left[\frac{\gamma_i\delta_{jk}+\gamma_j\delta_{ki}}{r} (1-p^2)
\right.
\nonumber\\&&
\left.
-\frac{2\gamma_i\gamma_j\gamma_k}{r} (1-p^2)
-\frac{\gamma_k\delta_{ij}}{r} (1+p^2)
\right]
\end{eqnarray}

The second derivatives are also obtained as
\begin{eqnarray}
\frac{\partial^2 G_{st,33}}{\partial\xi_i\partial x_j} 
({\bf x},\boldsymbol\xi)
&=&\frac{1}{2\pi\mu} \frac{\delta _{ij}-2\gamma_i\gamma_j}{r^2}
\\
\frac{\partial^2 G_{st,ij}}{\partial \xi_k\partial x_l} 
({\bf x},\boldsymbol\xi)
&=&-\frac{1}{4\pi\mu}
\left[
\frac{(\delta_{il}-2\gamma_i\gamma_l)\delta_{jk}}{r^2}(1-p^2)
\right.\nonumber\\&
+
&\left.
\frac{(\delta_{jl}-2\gamma_j\gamma_l)\delta_{ik}}{r^2}(1-p^2)
\right.\nonumber\\&
-
&\left.
\frac{(\delta_{kl}-2\gamma_k\gamma_l)\delta_{ij}}{r^2}(1+p^2)
\right.\nonumber\\&
-
&\left.
\frac{
\delta_{il}\gamma_j\gamma_k
+
\delta_{jl}\gamma_i\gamma_k
+
\delta_{kl}\gamma_i\gamma_j
-
4\gamma_i\gamma_j\gamma_k}{r^2}
\right.\nonumber\\&&\left.
\times2(1-p^2)
\right]
\end{eqnarray}
Note that the followings are useful;
$
\partial_i^{(x)} (\gamma_j/r)= (\delta_{ij}-2\gamma_i\gamma_j)/r^2
$,  
$\partial_i^{(x)} (\gamma_j\gamma_k)= 
(
\delta_{ij}\gamma_k+
\delta_{ik}\gamma_j-
2 \gamma_i\gamma_j\gamma_k
)/r^2
$.

\section{Numerical Precision of Piecewise-Constant Interpolation}
\label{sec:B}
Numerical precision of the line-chain result in \S\ref{sec:41} is explained analytically for the static two-dimensional problems. 
The original slip and axes of the local coordinates are supposed to be differentiable and the derivatives of them to be H\"{o}lder continuous. 
For simplicity, the slip gradient is supposed to exist in a finite region. 
Structured elements are treated mainly with the mid-point interpolation rule of the slip, and other cases are briefly mentioned. 

For brevity, we shorten the integral equation Eq.~(\ref{eq:stressGreenstatic}) to 
\begin{eqnarray}
\sigma_{kl}({\bf x}) &=& \int d\eta K_{kli}({\bf x};\boldsymbol\xi(\eta)) \partial^{(\xi)}_\tau \Delta u_i,
\\
K_{kli}({\bf x};\boldsymbol\xi)&=& 
c_{klmn}c_{ijpq}
 (\nu_j(\boldsymbol\xi)\tau_{m}(\boldsymbol\xi)-\tau_{m}(\boldsymbol\xi)\nu_j(\boldsymbol\xi)) 
\nonumber\\&&\times
\partial_q^{(\xi)} G_{st,np}({\bf x}-\boldsymbol\xi) 
\label{eq:shortenedkernel}
\end{eqnarray}
where $K$ denotes the shortened kernel, and $d\Sigma =d\eta$ is used.  
The static Green's function ${\bf G}_{st}({\bf x},\boldsymbol\xi)$ is rewritten as
${\bf G}_{st}({\bf x}-\boldsymbol\xi)$ given its spatial reciprocity.
% $\eta$ is a parameter continuous parameter on the fault to specify the fault position (as done in Fig. 1, left panel) and $K$ is a shortened kernel.
Hereinafter, we consider a given location ${\bf x}$ of the receiver as a fixed value, and then omit ${\bf x}$-dependence of $K$.
The location of the source $\boldsymbol\xi$ in the real space is parametrized by the local coordinate value $\eta$.

To begin with, the line-chain result is obtained 
by the following piecewise-constant interpolation of the slip in the global coordinate system:
\begin{eqnarray}
\Delta {\bf u}(\eta) \approx \sum_n \chi(\eta\in \Gamma_n) \Delta {\bf u}(\eta_n)
\label{eq:piecewisesliponsmooth}
\end{eqnarray}
or equivalently,
\begin{eqnarray}
\Delta {\bf u} \approx \sum_n \chi(\eta\in \Gamma_n) 
[\Delta u_\tau(\eta_n)\boldsymbol\tau(\eta_n)
+\Delta u_\nu(\eta_n)\boldsymbol\nu(\eta_n)],
\end{eqnarray}
where
 $\chi(\cdot)$ is the characteristic function that returns 1 when $(\cdot)$ is true and 0 otherwise, 
and 
$\Gamma_n$ denotes the area covered by the $n$-th discretized element. 
An inequality $\eta_{n+1}-\eta_n\leq\Delta \xi$ is satisfied for a given size $\Delta \xi$ of an element and gives $\eta_{n+1}-\eta_n=\mathcal O(\Delta \xi)$; the equality is achieved for the structured elements.  
The number of elements counted in the summation $\Sigma$ over the entire discretized area is of $\mathcal O((\Delta \xi)^{-1})$.

The stress integral is discretized into
\begin{eqnarray}
\int d\eta K_{abc} \partial_\tau^{(\xi)} \Delta u_c 
\approx
\sum_n
K_{abc}(\eta_{n}^+)  
[\Delta u_c(\eta_{n+1})
-
\Delta u_c(\eta_{n})]
\label{eq:midpointofutau}
\end{eqnarray}
or equivalently,
\begin{eqnarray}
&&\int d\eta K_{abc} \partial_\tau^{(\xi)} \Delta u_c 
\approx
\sum_n
K_{abc}(\eta_{n}^+)  
\nonumber\\&&
\,\,\,\,\,\,\,\,\,\,\,\,\,\,\,
\times\{
[\Delta u_\tau(\eta_{n+1})\tau_c(\eta_{n+1})
-
\Delta u_\tau(\eta_{n})\tau_c(\eta_{n})]
\nonumber\\&&
\,\,\,\,\,\,\,\,\,\,\,\,\,\,\,
+
[\Delta u_\nu(\eta_{n+1})\nu_c(\eta_{n+1})
-
\Delta u_\nu(\eta_{n})\nu_c(\eta_{n})]
\}
,
\label{eq:midpointofutaulocal}
\end{eqnarray}
where $\eta_n^+$ is the junction between the fault elements $n$, $n+1$, whose corresponding $\eta$ value in structure elements is the mean $\eta$ values over the fault elements $n$, $n+1$. 

Eq.~(\ref{eq:midpointofutaulocal}) is equivalent to the line-chain result in the case of the polygonal lines inscribed within the original curve. 
That indicates that Eq.~(\ref{eq:piecewisesliponsmooth}), the interpolation 
with respect to the slip, substantially includes the discretization of the fault geometry adopted in the line chain. 
Thus, the error estimate of the line-chain result is reduced to the error estimate of the piecewise constant slip on the smooth curve in our non-hypersingular expression.
Note that the higher-order interpolation requires to consider the boundary discretization independently from the interpolation of the slip, where the curvature arises explicitly.

\subsection{Accuracy of Off-Fault Stress}
Analytical proof is presented to the accuracy of the on-fault stress being the second order ($\mathcal O((\Delta \xi)^2)$) concerning the discretization width $\Delta \xi$. 

The order estimate of the accuracy is made of the following two steps.
First, the difference in $\Delta u_c$ in the right hand side of Eq.~(\ref{eq:midpointofutau}) is the mid-point interpolation of derivative of the slip in the global coordinate, 
as long as we use the structured elements, which satisfies $|\eta_n^+-\eta_n|=|\eta_n^+-\eta_{n+1}|$.
The error expressed by the second term is hence of $\mathcal O(\sum (\Delta \xi)^3)$;
\begin{eqnarray}
&&\sum_nK_{abc}(\eta_{n}^+)  
[\Delta u_c(\eta_{n+1})
-
\Delta u_c(\eta_{n})]
\nonumber\\
&&=
\sum_nK_{abc}(\eta_{n}^+)  
\left[
\partial^{(\xi)}_\tau 
(\Delta u_c)|_{\eta=\eta_n^+}
+ \mathcal O((\Delta \xi)^2)\right] \Delta \xi,
\label{eq:midpointofderiv}
\end{eqnarray}
Second, the leading order of the right hand side of Eq.~(\ref{eq:midpointofderiv}) is the mid-point interpolation of the integrand $K_{abc}\partial_\tau (\Delta u_c)$. Hence, as long as the $K_{abc}$ is regular for given $\eta$ (that is satisfied for the off-fault stress), the leading order
converges to the true solution up to the second order in each element. That is,
\begin{eqnarray}
&&
\sum_nK_{abc}(\eta_{n}^+)  
\left[
\partial_\tau^{(\xi)} 
(\Delta u_c)|_{\eta=\eta_n^+}
+ \mathcal O((\Delta \xi)^2)\right] \Delta \xi
\nonumber\\
&=&
 \int d\eta K_{abc} \partial_\tau^{(\xi)} \Delta u_c(1+\mathcal O((\Delta \xi)^2)) +\mathcal O(\sum (\Delta \xi)^3).
\\&=&
 \int d\eta K_{abc} \partial_\tau^{(\xi)} \Delta u_c +\mathcal O(\sum (\Delta \xi)^3).
\label{eq:suspiciousexp}
\end{eqnarray}
where the continuous integral is estimated at $\mathcal O(\int d\eta)=\mathcal(\sum\Delta \xi)$ in the second transform.
Eq.~(\ref{eq:suspiciousexp}) estimates the error to be of $\mathcal O(\sum (\Delta \xi)^3)$, which means $\mathcal O((\Delta \xi)^2)$. It is indeed what is observed in \S\ref{sec:41} numerically. 

Note that the error is expected to increase in the case of the unstructured elements or the interpolation other than the midpoint.
They deteriorate the error estimate in Eq.~(\ref{eq:midpointofderiv}) and 
$\mathcal O(\sum (\Delta \xi)^2)$ terms remain. The error is hence estimated to be of $\mathcal O(\Delta \xi)$ in the case of unstructured elements or non-midpoint interpolation. 

\subsection{Accuracy Deterioration of On-fault Stress}
Eq.~(\ref{eq:suspiciousexp}) contains expansion 
concerning $(\Delta \xi/|{\bf x}-\boldsymbol\xi|)$ due to that of the Green's function and its error scales as $\mathcal O((\Delta \xi)^2,(\Delta \xi/|{\bf x}-\boldsymbol\xi|)^2)$ more properly. 
Therefore, the above error estimate for the off-fault stress is not applicable to the on-fault stress, where the factor $(\Delta \xi/|{\bf x}-\boldsymbol\xi|)$ diverges at a point on the fault giving ${\bf x}-\boldsymbol\xi\to0$. 
The above estimate is modified below for the on-fault stress, numerically shown to contain the error of $\mathcal O(\Delta \xi)$ in \S\ref{sec:41}.

Without loss of generality, we suppose the receiver position at $\eta=0$ given the arbitrariness of the origin of the local coordinate system; we also assume that $\Delta \xi$ is normalized by some finite constant for brevity.
Next, we separate the slip in the region $|\xi|<M\Delta \xi$ around the receiver $\eta=0$ from the other (from $|\xi|>M\Delta \xi$) with a given small constant $M=o(1)$: 
\begin{eqnarray}
&&\int d\eta K_{abc} \partial^{(\xi)}_\tau \Delta u_c 
=\sigma_{near,ab}+\sigma_{dist,ab}
\label{eq:separatedone}
\\
&&\sigma_{near,ab}
:=
\int d\eta K_{abc} 
\nonumber\\&&\times
\partial^{(\xi)}_\tau\{ \Delta u_c [H(\eta+M\Delta \xi)-H(\eta-M\Delta \xi)]\}
\\
&&\sigma_{dist,ab}
:=
\int d\eta K_{abc} 
\nonumber\\&&\times
\partial^{(\xi)}_\tau\{ \Delta u_c [1-H(\eta+M\Delta \xi)+H(\eta-M\Delta \xi)]\}
\end{eqnarray} 
where $H(\cdot)$ is the Heaviside step function. 
The order of $M$ concerning $\Delta \xi$ is imposed arbitrarily, and is later set so as to get the correct order estimate of the error.

Below, we separately evaluate the error in Eq.~(\ref{eq:separatedone}) into those of $\sigma_{near,ab}$ for $|\eta|<M\Delta \xi$ and of $\sigma_{dist,ab}$ for $|\eta|>M\Delta \xi$.

We first evaluate the error in $\sigma_{near,ab}$.
In the case without discretization, the slip in the integrand of $\sigma_{near,ab}$ is expanded around the receiver while the kernel is kept unexpanded;
\begin{eqnarray}
&&\sigma_{near,ab}
\nonumber\\
&=&
\int d\eta K_{abc} \partial^{(\xi)}_\tau\{ (\Delta u_c(0)+\eta\Delta u_c^\prime(0)+\mathcal O(\eta^2) )
\nonumber\\&&\times
[H(\eta+M\Delta \xi)-H(\eta-M\Delta \xi)]\}
%\\
\end{eqnarray} 
\begin{eqnarray}
&=&
[K_{abc}(M\Delta \xi)-K_{abc}(-M\Delta \xi)]\Delta u_c(0)
\nonumber\\&&
+
[K_{abc}(M\Delta \xi)+K_{abc}(-M\Delta \xi)](\Delta u_c^\prime(0)M\Delta \xi)
\nonumber\\&&
+
\Delta u_c^\prime(0)\int^{M\Delta \xi}_{-M\Delta \xi} d\eta K_{abc}
+\mathcal O((M\Delta \xi)^2),
\label{eq:estimateofnearfault1}
\end{eqnarray} 
where $\Delta {\bf u}^\prime$ is the spatial derivative of the slip $\Delta {\bf u}$. 
We next utilize the symmetry of the homogeneous static Green's function:
\begin{eqnarray}
G_{st,ab}({\bf x}^\prime)=G_{st,ab}(-{\bf x}^\prime)
 \label{eq:symmetricforsourcereceiverlocation}
\end{eqnarray}
for arbitrary ${\bf x}^\prime$.
This symmetry is obtained from the equation of the stress balance: 
$ %\begin{eqnarray}
0= c_{ijpq}\partial^{(x)}_j\partial^{(x)}_p G_{st,qn}({\bf x}-\boldsymbol\xi)+\delta_{in}\delta ({\bf x}-\boldsymbol\xi),
%\label{eq:stressbalance}
$ %\end{eqnarray}
which is the quasi-static limit of the temporally integrated Eq.~(\ref{eq:EOMofG}).
%By comparing Eq.~(\ref{eq:stressbalance}) of (${\bf x}={\bf x}^\prime,\boldsymbol\xi=0$) with Eq.~(\ref{eq:stressbalance}) of (${\bf x}=0,\boldsymbol\xi={\bf x}^\prime$), 
%Eq.~(\ref{eq:symmetricforsourcereceiverlocation}) is obtained for 
%$G^{st}_{ab}({\bf x})=G^{hom,st}_{ab}({\bf x})$ due to $\delta({\bf x}^\prime)=\delta(-{\bf x}^\prime)$.
Given Eq.~(\ref{eq:symmetricforsourcereceiverlocation}) and
%the equalities, $G_{st,ab}({\bf x})=G_{st,ab}(-{\bf x})$ 
an equality $\partial^{(\xi)}_\tau (\tau_{j}\nu_m-\nu_m\tau_{j})=0$ for two-dimensional cases, 
we find  
the functional form, Eq.~(\ref{eq:shortenedkernel}), of the kernel 
is approximately antisymmetric for the on-fault receiver as 
\begin{eqnarray}
K_{abc}(\eta)=-K_{abc}(-\eta)+\mathcal O((M\Delta \xi)^2)
\label{eq:antisymmetryofkernelforsourcereceiverlocation}
\end{eqnarray}
 at $|\xi|\sim M\Delta \xi$, which is obtained with the expansion ${\boldsymbol\xi}(\eta)=$ $[d{\boldsymbol\xi}/d\eta(0)] \eta+...$. 
%This relation $K_{abc}(\eta)=-K_{abc}(-\eta)+\mathcal O((M\Delta \xi)^2)$
Eq.~(\ref{eq:antisymmetryofkernelforsourcereceiverlocation}) makes the second term of Eq.~(\ref{eq:estimateofnearfault1}) $\mathcal O((M\Delta \xi)^2)$;
\begin{eqnarray}
\sigma_{near,ab}
&=&
[K_{abc}(M\Delta \xi)-K_{abc}(-M\Delta \xi)]\Delta u_c(0)
\nonumber\\&&
+
\Delta u_c^\prime(0)\int^{M\Delta \xi}_{-M\Delta \xi} d\eta K_{abc}
+\mathcal O((M\Delta \xi)^2)
\label{eq:nearonfault}
\end{eqnarray} 

In the discretized case, the error in $\sigma_{near,ab}$ 
is evaluated with the expansion of the slip as 
\begin{eqnarray}
&&\mbox{discretized }
\sigma_{near,ab}
\nonumber\\
&=&
\sum_nK_{abc}(\eta_{n}^+)  
[\Delta u_c(\eta_{n+1})
-
\Delta u_c(\eta_{n})]
\nonumber\\&&
+K_{abc}(M\Delta \xi)\Delta u_c(M\Delta \xi+\Delta\xi/2)
\nonumber\\&&
-K_{abc}(-M\Delta \xi)\Delta u_c(-M\Delta \xi-\Delta\xi/2)
%\\&=&
%\sum_nK_{abc}(\eta_{n}^+) \Delta \xi [\Delta u_c^\prime(0)+\Delta u_c^{\prime\prime}(0)\eta+\mathcal O(\eta^2)]
%\nonumber\\
%&&+[K_{abc}(M\Delta \xi)-K_{abc}(-M\Delta \xi)]\Delta u_c(0)
%\nonumber\\
%&&+[K_{abc}(M\Delta \xi)+K_{abc}(-M\Delta \xi)]\Delta u^\prime_c(0)(M\Delta \xi+\Delta\xi/2)
%+\mathcal O((M\Delta \xi)^2).
\\&=&
[K_{abc}(M\Delta \xi)-K_{abc}(-M\Delta \xi)]\Delta u_c(0)
\nonumber\\&&
+\Delta u_c^\prime(0)
\int^{M\Delta \xi}_{-M\Delta\xi} d\eta K_{abc}(\eta_{n}^+)  
+\mathcal O((M\Delta \xi)^2).
\label{eq:nearonfaultdisc}
\end{eqnarray}
where the Taylor expansion of the slip gradient is used through the transform from the second to the third line with Eq.~(\ref{eq:antisymmetryofkernelforsourcereceiverlocation}).
%$K_{abc}(\eta)=-K_{abc}(-\eta)+\mathcal O((M\Delta \xi)^2)$. %is used from the second to the third, as in the un-discretized cases.

Comparing the expanded results of the un-discretized $\sigma_{near,ab}$
 and discretized one, respectively given in Eqs.~(\ref{eq:nearonfault}) and (\ref{eq:nearonfaultdisc}), 
the discretized error is noticed to be of $\mathcal O((M\Delta \xi)^2)$ for $\sigma_{near,ab}$:
\begin{eqnarray}
\sigma_{near,ab}-
\mbox{discretized }
\sigma_{near,ab}=\mathcal O((M\Delta \xi)^2).
\label{eq:onfaultdiscerrornear}
\end{eqnarray}

Meanwhile, the error in $\sigma_{dist,ab}$ is of $\mathcal O((\Delta \xi)^2,1/M^2)$ given the same expansion as for the off-fault stress, 
\begin{eqnarray}
\sigma_{dist,ab}-
\mbox{discretized }
\sigma_{dist,ab}=\mathcal O((\Delta \xi)^2,1/M^2)
\label{eq:onfaultdiscerrordist}
\end{eqnarray}
by considering that the error in Eq.~(\ref{eq:suspiciousexp}) is ($\mathcal O((\Delta \xi)^2,(\Delta\xi/|{\bf x}-\boldsymbol\xi|)^2)$) due to the Taylor expansion on $\Delta\xi/|{\bf x}-\boldsymbol\xi|$. 

%the errors of $others$ can be evaluated at $\mathcal O(\sum (\Delta \xi)^2)$ as in the off-fault stress, as long as $M\Delta \xi$ is sufficiently small so as to make the minimum of $|\xi(\eta)|$ be of $\mathcal O(M\Delta \xi)$ for $|\eta|>M\Delta$; this relation is satisfied by the sufficiently narrow boundary range compared with the curvature radius of the fault.

The lower bound of the above error estimate is obtained with $M$ that matches errors in $\sigma_{near,ab}$ [Eq.~(\ref{eq:onfaultdiscerrornear})] and $\sigma_{dist,ab}$, [Eq.~(\ref{eq:onfaultdiscerrordist})], i.e., $(M\Delta \xi)^2\sim \max (1/M^2,(\Delta \xi)^2)$. 
This $M$ value, $M=\mathcal O(\sqrt{\Delta \xi})$, predicts that the error is of $\mathcal O(\Delta \xi)$ in total, 
%Therefore, the error at $|\xi|<M\Delta \xi$ becomes of $\mathcal O((M\Delta \xi)^2)$, and the total error becomes of $\mathcal O(\Delta \xi)$ (since $M$ is chosen to match the errors around the receiver and of the others as $(M\Delta \xi)^2\sim 1/M^2$). 
which is consistent with our numerical results (Fig.~\ref{fig:2}, bottom).

The error estimate of $\sigma_{near}$ does not rely on neither the midpoint interpolation rule nor structured elements. 
The error becomes of $\mathcal O(M^{-1},\Delta \xi)$ for $\sigma_{dist}$ if either of these conditions is not valid, according to the similar logic to the off-fault one. Given these, the error of the on-fault stress will become of $\mathcal O((\Delta \xi)^{2/3})$ with $M=\mathcal O((\Delta \xi)^{2/3})$ [that satisfies $(M\Delta \xi)^2\sim \max(1/M,\Delta \xi)$] for non-midpoint interpolations or unstructured elements.

\section{Comparison of the nomenclature for the non-hypersingular expressions}
\label{sec:comparisonofnomenclature}
\renewcommand{\thetable}{C.\arabic{table}}
\begin{table*}
\begin{tabular}{l|cccc}
\hline
& This Study
& Group A
& Companion Study
&  Group B
\\
\hline
Location of Receiver
&${\bf x}$
&${\bf x}$
&${\bf x}$
&${\bf x}$
\\
Time of Receiver 
& $t$
& $t$
& $t$
& $t$
\\
Source Location in Global Coordinates
&$\boldsymbol\xi$
&$\boldsymbol\xi$
&${\bf y}$
&${\bf y}$
\\
Source Location in Local Coordinates 
&$\boldsymbol\eta$
&
&
&s
\\
Time of Source
&$s$
&$\tau$
&$\tau$
&$\tau$
\\
Normal Vector
&$\boldsymbol\nu$
&$\boldsymbol\nu$
&${\bf n}$
&${\bf n}$
\\
Tangential Vectors
&$\boldsymbol\tau_1, \boldsymbol\tau_2$
&
&${\bf t},{\bf s}$
&${\bf t},{\bf s}$
\\
Subscripts 
&
$m,n,i,j,p,q$
&
$n,i,j,p,q$
&
$c,d,i,j,p,q$
&
$k,l,i,j,p,q$
\\
\hline
\end{tabular}
\caption{Comparison table of symbols, between 
this study, 
group A following \citet{aki2002quantitative}, 
the companion paper of this study (companion study), and 
group B following \citet{tada1997non}. 
Blank represents the lack of symbols.
For the group B, we also referred to \citet{tada1996paradox} for the local coordinate values and 
\citet{tada2000non} for three-dimensional cases. 
}
\label{tab:symbols}
\end{table*}

Non-hypersingular stress Green's function specifies a number of variables, such as 1) location of receiver, 2) time of receiver, 3) location of source in the global coordinates, 4) time of source, 5) normal vector on the fault, and 6) tangential vectors on the fault. Furthermore, since the local coordinate system is curvelinear, we need to distinguish 7) the location of the source in the local coordinates from that in the global coordinates, to relate the differentials in the (two-dimensional curvelinear) local coordinates with that of (three-dimensional Euclidean) global coordinates. The nomenclature of the non-hypersingular expressions is then complicated. We list them in Table~\ref{tab:symbols}.

Nomenclatures of the previous studies are mostly separated into a group A following \citet{aki2002quantitative} and B following \citet{tada1997non}. 
We followed the group A unless the duplication of symbols arises. 
Please refer to the companion paper (Romanet et al.) for the non-hypersingular expression in the nomenclature of the group B.
Note that in the other studies, readers need to pay attention to the point that the subscripts $t$ to express a component parallel to the tangential vector can be duplicated with time $t$ of receiver.
Duplication of symbols is not contained in this paper.

\end{document}